\documentclass[10pt,aps,prd,superscriptaddress,nofootinbib,nobibnotes,longbibliography,floatfix,twocolumn]{revtex4-2}

\usepackage{bm}
\usepackage{mathtools,
amsmath,
amssymb,
amsfonts,
mathrsfs,
chngcntr,
multirow}

\let\cc\corresponds
\let\corresponds\relax
\usepackage{mathabx}
\let\corresponds\cc

\usepackage[utf8]{inputenc}
\usepackage[T1]{fontenc}

\usepackage[dvipsnames]{xcolor}
\usepackage[unicode]{hyperref}
\hypersetup{colorlinks=true, citecolor=MidnightBlue,
            linkcolor=MidnightBlue, urlcolor=MidnightBlue, linktocpage=true}
\usepackage[normalem]{ulem}

\DeclareMathAlphabet{\mathpzc}{OT1}{pzc}{m}{it}
\definecolor{darkgreen}{rgb}{0.0, 0.6, 0.0}

\newcommand{\ie}{\textit{i.e.}}
\newcommand{\eg}{\textit{e.g.}}
\newcommand{\note}[1]{\text{\scshape\tiny{#1}}}

\newcommand{\ee}{\mathrm{e}}
\newcommand{\ii}{\mathrm{i}}
\newcommand{\dd}{\mathrm{d}}

\newcommand{\GN}{G_\note{N}}

\newcommand{\Ord}{\mathcal{O}}

\newcommand{\al}{\alpha}
\newcommand{\be}{\beta}
\newcommand{\ga}{\gamma}

\newcommand{\de}{\delta}
\newcommand{\De}{\Delta}

\newcommand{\cep}{\varepsilon}

\renewcommand\th{\theta}

\newcommand{\la}{\lambda}
\newcommand{\La}{\Lambda}
\newcommand{\sg}{\sigma}

\newcommand{\om}{\omega}

\newcommand{\dl}{\partial}

\newcommand{\tcb}[1]{\textcolor{black}{#1}}

\begin{document}
\title{Parametrized quasi-normal mode framework for modified Teukolsky equations}

\author{Pablo A. Cano}
\email{pablo.cano@icc.ub.edu}
\affiliation{Departament de F\'isica Qu\`antica i Astrof\'isica, Institut de Ci\`encies del Cosmos\\
 Universitat de Barcelona, Mart\'i i Franqu\`es 1, E-08028 Barcelona, Spain}
 
 \author{Lodovico Capuano}
\email{lcapuano@sissa.it}
\affiliation{SISSA, Via Bonomea 265, 34136 Trieste, Italy and INFN Sezione di Trieste}
\affiliation{IFPU - Institute for Fundamental Physics of the Universe, Via Beirut 2, 34014 Trieste, Italy}

 \author{Nicola Franchini}
\email{franchini@apc.in2p3.fr}
\affiliation{Universit\'e Paris Cit\'e, CNRS, Astroparticule et Cosmologie, F-75013 Paris, France}
\affiliation{CNRS-UCB International Research Laboratory, Centre Pierre Bin\'etruy,
IRL2007, CPB-IN2P3, Berkeley, CA 94720, USA}

 \author{Simon Maenaut}
\email{simon.maenaut@kuleuven.be}
\affiliation{Institute for Theoretical Physics, KU Leuven. Celestijnenlaan 200D, B-3001 Leuven, Belgium}
\affiliation{Leuven Gravity Institute, KU Leuven. Celestijnenlaan 200D, B-3001 Leuven, Belgium}

 \author{Sebastian H. V\"olkel}
\email{sebastian.voelkel@aei.mpg.de}
\affiliation{Max Planck Institute for Gravitational Physics (Albert Einstein Institute),
D-14476 Potsdam, Germany}

\begin{abstract}
Modifications to general relativity lead to effects in the spectrum of quasi-normal modes of black holes. In this paper, we develop a parametrized formalism to describe deviations from general relativity in the Teukolsky equation, which governs linear perturbations of spinning black holes. We do this by introducing a correction to the effective potential of the Teukolsky equation in the form of a $1/r$ expansion controlled by free parameters. The method assumes that a small deviation in the effective potential induces a small modification in the spectrum of modes and in the angular separation constants. We isolate and compute the universal linear contribution to the quasi-normal mode frequencies and separation constants in a set of coefficients, and test them against known examples in the literature (massive scalar field, Dudley-Finley equation and higher-derivative gravity). We make the coefficients publicly available for relevant overtone, angular momentum and azimuthal numbers of modes and different values of the black hole spin.
\end{abstract}

\maketitle

\section{Introduction}

Gravitational wave astronomy successfully observed more than one hundred binary black hole (BH) mergers~\cite{KAGRA:2021vkt}. Such an advancement in the field allows one to make precision tests of general relativity (GR). In particular, BH spectroscopy, \ie, the identification of the \tcb{different quasi-normal modes} (QNMs) of which the signal is composed in the linear post-merger phase known as the ringdown~\cite{Kokkotas:1999bd,Berti:2009kk,Franchini:2023eda}, has finally been applied to tens of events~\cite{LIGOScientific:2016lio,LIGOScientific:2020tif,LIGOScientific:2021sio}. The detection of two modes simultaneously, despite being controversial for the first three observing runs~\cite{Carullo:2019flw,Isi:2019aib,Cotesta:2022pci,Finch:2022ynt,Isi:2022mhy,Carullo:2023gtf,Crisostomi:2023tle,Pacilio:2024qcq,Gennari:2023gmx}, is expected to be effective for the current O4a run and future ones.

Detecting a second mode is a crucial ingredient for tests of GR. Due to no-hair theorems, the QNM spectrum of a Kerr BH depends uniquely on its mass and spin. If only one mode is detected, this can always be fitted to a QNM frequency of a Kerr BH with a certain mass and angular momentum. However, the measurement of any additional modes provides a consistency test of the Kerr QNM spectrum and hence would allow us to spot deviations from GR. 
This clear identification of possible beyond-GR effects makes BH spectroscopy one of the most promising ways to test GR.

Currently, ringdown tests employ blind deviations from GR in the frequencies~\cite{LIGOScientific:2021sio}, or agnostic deviations constructed assuming small-coupling and slow-spin parametrization~\cite{Maselli:2019mjd,Carullo:2021dui,Maselli:2023khq}. On the other hand, theory-specific tests are limited to a handful of cases~\cite{Carullo:2021oxn}. This is because the computation of QNMs for rotating solutions beyond-GR is incomplete. The main difficulties arise from: absence of analytic background solution, non-separability of the perturbation equations, additional fields coupled to the metric, different boundary conditions~\cite{Franchini:2023eda}. It turns out that for a vast class of theories, the first problem can be solved performing a double simultaneous expansion in the spin and in the coupling constant of the theory \cite{Pani:2009wy,Yunes:2009hc,Pani:2011gy,Maselli:2015tta,Cano:2019ore}. Then, one can choose whether to study perturbations giving priority to slow-spin or small coupling. The former has the advantage of being possible for metric perturbations, for which couplings between different fields are tractable, at the cost of not being able to predict precisely QNMs at high spins (which are relevant for astrophysical purposes)~\cite{Pani:2013pma,Franchini:2023xhd,Cano:2021myl,Pierini:2021jxd,Pierini:2022eim,Wagle:2021tam,Srivastava:2021imr}.

On the other hand, by assuming small-coupling for the perturbations, one can work out a modified Teukolsky equation, which is, in principle, reliable at any spin~\cite{Li:2022pcy,Hussain:2022ins,Cano:2023tmv}. The disadvantage comes from the construction itself of the Teukolsky equation, which is based on curvature perturbations, and once one has some perturbations of perturbations, as in the case outlined here, metric reconstruction becomes necessary. This feature strongly hinders one from going beyond the first order in the coupling expansion.

The necessity of having reliable QNMs at high spins and the fact that observations seem to narrow down the size of deviations from GR, make the modified Teukolsky framework preferable for the study of QNMs beyond GR. The scope of this paper is to develop a general formalism for the quick computation of QNMs in theories which have a perturbative departure from GR. The framework is based on an assumption similar to one developed in spherical symmetry~\cite{Cardoso:2019mqo,McManus:2019ulj,Volkel:2022aca,Hirano:2024fgp}. The advantage of this formalism is that it can be used also for the inverse problem, \ie, if a modification to GR is detected, one wants to be able to reconstruct the potential, the metric, or even the action from which the deviation originated from~\cite{Volkel:2022aca,Volkel:2022khh,Franchini:2022axs}. In general, this framework opens the path to the development of a theory-informed description of QNM agnostic deviations.

The structure of the paper is as follows: we first introduce the modified Teukolsky equation and the formalism in section~\ref{sec:param}; then we show how to numerically compute the coefficients with the continued fraction method and their regime of validity in section~\ref{sec:continued_fraction}; the formalism has already a few notable applications, which we use as a check for our computations in section~\ref{sec:applications}; finally, we outline our conclusions in section~\ref{sec:conlusions}. Throughout the paper we work with mostly plus signature for the metric, and geometric units $\GN = c = 1$. We also assume that the background, unmodified metric is a Kerr BH of mass $M=1/2$ and spin $a$. It is worth noting that due to the choice of units, the spin parameter $a$ ranges between $0$ and $1/2$.

\section{Parametrized formalism}\label{sec:param}

\subsection{The linear coefficients}

Let us start from the radial Teukolsky equation for a spin $s$ field as~\cite{Teukolsky:1973ha}
\begin{equation}\label{eq:Teukolsky}
    \frac{1}{\De^s R(r)} \frac{\dd}{\dd r} \left[ \De^{s+1} R'(r) \right] + V(r) = 0 \,,
\end{equation}
where the effective potential reads
\begin{equation}
    V(r) = 2\ii s \frac{\dd K}{\dd r} - \la_{\ell m} + \frac{1}{\De} \left( K^2 - \ii s K \frac{\dd \De}{\dd r}\right) \,,
    \label{Teukolsky_potential}
\end{equation}
and we defined
\begin{align}
    \De & = r^2 - r +a^2 \,, \qquad K = (r^2+a^2)\om - a m \,, \\
    \la_{\ell m} & = B_{\ell m} +a^2\om^2 - 2 a m \om \,.
\end{align}
It is worth mentioning that the zeros of the function $\De$ determine the location of the BH inner and outer horizons, given by
\begin{equation}
  r_\pm = \frac{1 \pm \sqrt{1-4a^2}}{2} \,.  
\end{equation}
On the other hand, the angular equation reads~\cite{Teukolsky:1973ha}
\begin{equation}\label{eq:Teuk_ang}
\begin{split}
    \frac{1}{S(y)} & \frac{\dd}{\dd y} \left[ (1-y^2) S'(y) \right] + a^2 \om^2 y^2 \\
    & - 2s a \om y +B_{\ell m} + s - \frac{(m + s y)^2}{1-y^2}  = 0 \,,
\end{split}
\end{equation}
where $y = \cos\th$.

We now assume that for a modified theory of gravity whose modifications are small with respect to GR the equation governing radial perturbations is the Teukolsky radial equation plus a correction to the potential linear in the coupling constants
\begin{equation}\label{eq:Teukolskymod}
    \frac{1}{\De^s R(r)} \frac{\dd}{\dd r} \left[ \De^{s+1} R'(r) \right] + V(r) + \de V(r)  = 0 \,,
\end{equation}
and we assume that the modification is expanded in powers of $r$
\begin{equation}\label{eq:delta_potential}
    \de V(r) = \frac{1}{\De}\sum_{k= - K}^{4} \al^{(k)} \left(\frac{r}{r_+}\right)^k\,,
\end{equation}
where $K$ is \tcb{the most negative coefficient} of the power series and $\al^{(k)}$ are dimensionful coefficients we assume to be small. This assumption is justified by the recent developments in obtaining modified Teukolsky equations by assuming small coupling corrections to GR~\cite{Li:2022pcy,Hussain:2022ins,Cano:2023tmv}.

On the other hand, we can also assume that the angular equation remains unchanged. This is due to the fact that the spheroidal harmonics are a complete basis of the 2-sphere angular variables, and it can be shown that if one forces an angular expansion of the Weyl scalars in spin-weighted spheroidal harmonics, all the mixing terms would enter at second order in the coupling constants~\cite{Cano:2023tmv,Ghosh:2023etd}. Nevertheless, a modification on the QNM frequencies will induce a modification to the separation constant $B_{\ell m}$, hence, we also need to include equation~\eqref{eq:Teuk_ang} in our analysis.

If we assume the couplings to be small, \footnote{\tcb{We will provide a more quantitative comment about this assumption in Sec. \ref{Linearized_regime}. For now just consider these dimensionless parameters to be small enough to allow a linearization of the eigenvalues $\omega_{n\ell m}$ and $B_{\ell m}$.}} we are allowed to perform a Taylor expansion of QNMs and separation constants around their GR values \cite{Cardoso:2019mqo,McManus:2019ulj}. Hence, we can write 
\begin{equation}\label{eq:dOmega_dB}
\begin{split}
    \omega_{n\ell m} & \simeq \omega_{n\ell m}^0 + \sum_{k} d_{\om, n\ell m}^{(k)} \al^{(k)} \,, \\
    B_{\ell m}(a\om) & \simeq B_{\ell m}^0(a\om) + \sum_{k} d_{B, \ell m}^{(k)} \al^{(k)} \,.
\end{split}
\end{equation}
In the following steps, we omit the indices $s, n, \ell, m, a, \om$ for clarity. The linear coefficients $d_\om$ and $d_B$ can be identified as the derivatives of $\omega$ and $B$ with respect to the single coupling $\alpha$. To compute them, we perform the following steps. In the GR limit, one finds $\om$ and $B_{\ell m}$ as simultaneous roots of two functions constructed from the radial and the angular equation
\begin{align}
    \mathcal{L}_r\left[\om,B_{\ell m} \right] = 0 \,, \qquad \mathcal{L}_\th\left[\om,B_{\ell m} \right] = 0 \,.
    \label{Leaver_eq}
\end{align}
The exact form of these functions depends on the chosen numerical method. For non-zero modifications, we can perform a Taylor expansion of the two functions around $\al = 0$
\begin{equation}\label{eq:Taylor}
    \left.\mathcal{L}_j\right|_{\note{GR}} + \al \left.\frac{\dd \mathcal{L}_j}{\dd \al}\right|_\note{GR}  + \Ord(\al)^2 = 0\,,
\end{equation}
where $j=[r,\th]$ and we evaluate the derivative around their GR value $\omega=\omega^0$, $B = B^0$, $\alpha = 0$. By requiring that equations~\eqref{eq:Taylor} is satisfied at each order in $\al$, and expanding the derivative by chain rule, we obtain
\begin{equation}
\begin{split}\label{eq:r_th_chain}
    \left.\frac{\partial\mathcal{L}_r}{\partial\alpha}+\frac{\partial\mathcal{L}_r}{\partial\omega}\,d_\omega+\frac{\partial\mathcal{L}_r}{\partial B}\,d_B \right|_\note{GR} &= 0 \,, \\
    \left.\frac{\partial\mathcal{L}_\theta}{\partial\omega}\,d_\omega+\frac{\partial\mathcal{L}_\theta}{\partial B}\,d_B \right|_\note{GR} &= 0\,,
\end{split}
\end{equation}
where we identified $d_\om$ and $d_B$ from their definition in equation~\eqref{eq:dOmega_dB}.
By solving the conditions above for $d_\omega$ and $d_B$, we get
\begin{equation}\label{eq:coefficients_d}
\begin{split}
    &d_\omega = \left.-\frac{\dl\mathcal{L}_r}{\dl \alpha}\,\frac{\dl\mathcal{L}_\theta}{\dl B}\,\left(\frac{\dl\mathcal{L}_r}{\dl \omega}\frac{\dl\mathcal{L}_\th}{\dl B}-\frac{\dl\mathcal{L}_r}{\dl B}\frac{\dl\mathcal{L}_\th}{\dl \om}\right)^{-1} \right|_\note{GR} \,, \\
    &d_B = \left. \frac{\dl\mathcal{L}_r}{\dl \alpha}\,\frac{\dl\mathcal{L}_\theta}{\dl \omega}\,\left(\frac{\dl\mathcal{L}_r}{\dl \omega}\frac{\dl\mathcal{L}_\th}{\dl B}-\frac{\dl\mathcal{L}_r}{\dl B}\frac{\dl\mathcal{L}_\th}{\dl \om}\right)^{-1} \right|_\note{GR} \,.
\end{split}
\end{equation}
In section~\ref{sec:continued_fraction} we show how to numerically define the functions $\mathcal{L}_j$ with Leaver's continued fraction method.

\subsection{Maximum number of independent coefficients}\label{sec:ambiguity}

In reference~\cite{Kimura:2020mrh}, Kimura realised that in the case of spherically symmetric perturbations, there is always an ambiguity in defining the modified potential, upon a free reparametrization of the field. The same reasoning can be applied to the Teukolsky equation as well. If we perform the following transformation in equation~\eqref{eq:Teukolskymod}
\begin{equation}\label{eq:transformation_ambiguous}
    R(r) \rightarrow \left[1 + \cep X(r) \right] R(r) + \cep \De Y(r) R'(r) \,,
\end{equation}
assuming that $\cep \ll 1$, then the equation that $R(r)$ solves is
\begin{equation}\label{eq:Teukolskymod2}
    \frac{1}{\De^s R} \frac{\dd}{\dd r} \left[ \De^{s+1} R' \right] + V + \de V + \de \overline{V} + \de W \frac{R'}{R}  = 0 \,.
\end{equation}
By imposing $\de W = 0$ we uniquely obtain the free function $X(r)$ as
\begin{equation}
    X(r) = c + \frac{s}{2} \De' Y - \frac{1}{2} \De Y'\,,
\end{equation}
which yields the "ambiguous" potential in the form
\begin{equation}
\begin{split}
    \de \overline{V} = \cep \De \Bigg[ & Y' \left(\frac{\left(s^2-1\right) \left(\Delta '\right)^2}{2 \Delta }+2 s-2 V-1\right) \\
    & +Y \left(\frac{s (s+1) \Delta '}{\Delta }-V'-\frac{V \Delta '}{\Delta
   }\right) \\
   & -\frac{1}{2} \Delta  Y^{(3)}-\frac{3 \Delta ' Y''}{2} \Bigg] \,.
\end{split}
\end{equation}
Now, upon suitable choice of the function $Y$, we can express $\de \overline{V}$ in the $r$ basis. It turns out that the ansatz $Y = Y_j = y_j (r_+/r)^j$ yields
\begin{equation}\label{eq:ambiguous}
\begin{split}
    \de \overline{V} = \frac{\cep y_j}{\De} \sum_{k=-3}^{5} r_+^k \overline{A}_j^{(k)} \left( \frac{r}{r_+} \right)^{k-j} \,,
\end{split}
\end{equation}
which implies that $j \geq 1$ since the maximum power of $r$ in $V$ is $r^4$, and the full expression of $\overline{A}_j^{(k)}$ can be found in appendix~\ref{app:ambiguity}. In general, one can take a linear combination of the free functions $Y_j$ and still get a potential that is equivalent to the starting one. Each term of this linear combination contains the free parameter $y_j$, which can be used to set to $0$ one of the terms $\al^{(k)}$ in equation~\eqref{eq:delta_potential}. This reasoning allows us to fix the negative limit in the power expansion to be $K = 3$.

It is possible that by choosing a different ansatz for $Y$ one could further reduce the number of coefficients in the equation. In fact, for the case of study of higher derivative gravity that we treat in another publication~\cite{Cano:2024ezp} the number of independent coefficients reduces to four (being $k=[-2,0,1,2]$) --- see also \cite{Cano:2023jbk}.
Although we could not prove this is a general feature of arbitrary modifications of the Teukolsky equation, we suspect that it was possible in that case thanks to the expansion in the spin assumed for every coefficient. Indeed, we believe that the ansatz for $Y$ that would reduce the potential to the lowest number of terms would be, perhaps, a rational function involving powers of $a$ and $r$. To date, we could not find such reduction.

\section{Computation of the coefficients: the continued fraction method}\label{sec:continued_fraction}

\subsection{Continued fractions for the Teukolsky equation}

We start here by recalling the Leaver method to compute the frequencies and the separation constant for a Kerr spacetime. The first step to find a continued fraction expansion is to assume an ansatz for the wavefunctions. Let us start from the radial equation, where we assume the following ansatz~\cite{Leaver:1985ax}
\begin{equation}\label{ansatz}
     R(r) = f^{-\ii \sg - s} (r- r_-)^{p -1 -2s} \ee^{q r} \sum_{n = 0}^N R_n f^n\,,
\end{equation}
where $r_\pm$ are the zeros of $\De$,
\begin{equation}
    f = \frac{r - r_+}{r - r_-}\,,
\end{equation}
and we defined $p=q=\ii\om$ and
\begin{align}
    & \sg = \sg_\note{GR} \equiv \frac{r_+ ( \om - \om_c)}{r_+ - r_-}\,, \qquad  \om_c = \frac{a m}{r_+}\,, \\
    & r_\pm = \frac{1}{2}(1\pm\be) \,, \qquad \be = \sqrt{1-4a^2} \,.
\end{align}
With these definitions, the equation~\eqref{eq:Teukolsky} takes the form
\begin{equation}\label{eq:Leaver_Teuk}
    \sum_{n = 0}^N R_n \left( \frac{\al^r_{n-1}}{f} + \be^r_{n} + \ga^r_{n+1} f \right) f^n = 0 \,,
\end{equation}
where the coefficients are
\begin{align}
    \al^r_n = & \, (n+1) \left( n + 1 - s -2\ii \sg \right) \\
    \be^r_n = & \, 2n(2\ii \sg + p + q \be -1) - 2n^2 - 1 - s - B_{\ell m} \notag \\
    & +q \left(a^2 q+\be +s\right)-2 i \sg  (p+\be  q+i \om -1) \notag \\
    & -p (\be  q+q+s-1)\\
    \ga^r_n = & \, \left(n - p - \ii \om \right) \left(n + s - p - 2\ii\sg + \ii\om \right) \,.
\end{align}
The equation is satisfied when each term proportional to a power of $f$ vanishes
\begin{align}\label{eq:coeff_0relation}
    \be^r_0 R_0 + \al^r_0 R_1 & = 0\,, \\
    \ga^r_{n} R_{n-1} + \be^r_{n} R_{n} + \al^r_{n} R _{n+1} & = 0 \quad \text{for } n \geq 1\,.   \label{eq:coeff_relation}
\end{align}
The path for the angular equation is similar. We define an ansatz to be finite at the regular singular points $y = \pm 1$~\cite{Leaver:1985ax}
\begin{equation}
    S(y) =  (1+y)^{k_1} (1-y)^{k_2} \ee^{a \om y} \sum_{n=0}^N S_n (1+y)^n\,,
\end{equation}
where $k_1 = |m-s|/2$ and $k_2 = |m+s|/2$. We can obtain a similar recurrence relation by inserting this ansatz into equation~\eqref{eq:Teuk_ang}, which, with an analogous reshuffling, reads
\begin{align}\label{eq:coeff_S0relation}
    \be^\th_0 S_0 + \al^\th_0 S_1 & = 0\,, \\
    \ga^\th_{n} S_{n-1} + \be^\th_{n} S_{n} + \al^\th_{n} S _{n+1} & = 0 \quad \text{for } n \geq 1\,,
\end{align}
where the coefficients are
\begin{align}
    \al^\th_n = & \, - 2 (n+1) \left(n + 1 + 2 k_1 \right)\,, \label{eq:coeff_th_1} \\
    \be^\th_n = & \, n(n-1) + 2n (k_1 + k_2 + 1 - 2 a \om) \notag \\
    & - 2 a \om( 2 k_1 + s +1) + (k_1 + k_2) (k_1 + k_2 + 1 ) \notag \\
    & - a^2 \om^2 - s(s+1) - B_{\ell m}\,, \label{eq:coeff_th_2} \\
    \ga^\th_n = & \, 2 a \om (n + k_1 + k_2 + s)\,. \label{eq:coeff_th_3}
\end{align}

To invert the relation we can define the ladder operators which have the following property $R_{n+1} = -\La^r_n R_n$ and $S_{n+1} = -\La^\th_n S_n$ as (the superscript $r/\th$ is omitted for clarity)
\begin{equation}\label{eq:cf_relation}
    \La_n = \frac{\ga_{n+1}}{\be_{n+1} - \al_{n+1} \La_{n+1}}\,.
\end{equation}
By initializing $\La_N$ according to the Nollert expansion (explained in detail in appendix~\ref{app:Nollert}), the equations one needs to solve simultaneously to obtain the eigenfrequency $\om$ and the separation constant $B_{\ell m}$ are
\begin{align}
    \mathcal{L}_r = \La_1^{r} \al^{r}_0 - \be^{r}_0 & = 0 \, , \label{eq:Leaver_condition_r} \\
    \mathcal{L}_\th = \La_1^{\th} \al^{\th}_0 - \be^{\th}_0 & = 0 \, ,\label{eq:Leaver_condition_th} 
\end{align}
which are nothing but \eqref{eq:coeff_0relation} and \eqref{eq:coeff_S0relation}.

\subsection{Continued fraction beyond Teukolsky}

We now turn our attention to the modified Teukolsky equation. It is always possible to bring equation~\eqref{eq:delta_potential} into the following form\footnote{In section~\ref{sec:ambiguity} we showed that $K=3$, but the following analysis works, in principle, for any value of $K$, hence we keep it unspecified.}
\begin{equation}\label{eq:delta_potential_2}
\begin{split}
    \de V(r) = & \, \frac{A^{(0)}}{\De} + \frac{A^{(1)}}{r_+(r-r_-)} +  \frac{1}{r_+^2}\sum_{k = 0}^2 \widetilde{\al}^{(k)} \left(\frac{r}{r_+}\right)^k \\
    &  + \frac{1}{\De}\sum_{k = 1}^K \al^{(-k)} \left(\frac{r_+}{r}\right)^k\,,
\end{split}
\end{equation}
where $A^{(0)}$, $A^{(1)}$ and $\widetilde{\al}^{(k)}$ are constants that can be obtained from the constants $\al^{(k)}$ appearing in equation~\eqref{eq:delta_potential}, as explained in appendix~\ref{app:potential}. First of all, we notice that the terms multiplied by $1/\De$ modify the behaviour of the equation at the horizon. In order to take into account of these additional terms, we need to modify the definition of the exponent $\sg$ appearing in the ansatz~\eqref{ansatz}. By requesting that the solution is regular at the horizon, we must replace the value of $\sg$ into
\begin{equation}\label{eq:hor_expansion}
    \sg = \frac{\ii s}{2} + \sqrt{\left(\sg_\note{GR} - \frac{\ii s}{2} \right)^2 + \frac{1}{\be^2}\sum_{k=-K}^4 \al^{(k)} } \,,
\end{equation}
where we took the positive sign of the square root in order to obtain the correct GR limit. 
On the other hand, the terms $\widetilde{\al}^{(1)}$ and $\widetilde{\al}^{(2)}$ modify the behaviour at infinity of the equation. This leads to a modification of the values of $p$ and $q$ into
\begin{align}
    \label{eq:infty_q}
    & q=\pm\sqrt{-\frac{\widetilde{\al}^{(2)}}{r_+^4}-\om^2}\,, \\
    \label{eq:infty_p}
    & p=-\frac{r_+\widetilde{\al}^{(1)} + \widetilde{\al}^{(2)} - 2 r_+^4 \left( q s - \ii s\om - \om^2 \right)}{2q r_+^4}\,,
\end{align}
where the sign of $q$ is chosen such that $\text{Re}(q) > 0$. \tcb{See Appendix \ref{BCs} for the derivation of the appropriate boundary conditions for the beyond-Teukolsky case.}
This asymptotic behaviour is the reason why we truncate the series in equation~\eqref{eq:delta_potential} at $k=4$. 
By repeating the steps done for the GR case, we obtain a modified version of equation~\eqref{eq:Leaver_Teuk}
\begin{equation}\label{eq:Leaver_Teuk_mod}
\begin{split}
    \sum_{n = 0}^N & R_n \Bigg[ \frac{\al^\note{bg}_{n-1} }{f} + \be^\note{bg}_{n}  + \ga_{n+1}^\note{bg} f \\
    & + \frac{1}{\be^2}\frac{(1-f)^2}{f} \sum_{k=1}^{K} \al^{(-k)} \left( \frac{1- f}{ 1 - \eta f} \right)^k \Bigg] f^n = 0\,,
\end{split}
\end{equation}
where $\eta = r_- / r_+ $ and
\begin{align}
    \al_{n}^\note{bg} = & \, \al^r_{n} - \frac{1}{\be^2}\sum_{k=1}^K \al^{(-k)}\,, \\
    \be_{n}^\note{bg} = & \, \be^r_{n} - 2\left(\sg_\note{GR} - \sg \right) \left(\sg_\note{GR} + \sg - \om \right)\,, \notag \\
    & -2 \frac{A^{(0)}}{\be^2} + \frac{A^{(1)}}{r_+\be} + \frac{\widetilde{\al}^{(0)}}{r_+^2} \\
    \ga_{n}^\note{bg} = & \, \ga^r_{n} + 2\ii\left(\sg_\note{GR} - \sg \right)(s+\ii\om) \notag \\
    & - \frac{1}{\be^2}\sum_{k=1}^K \al^{(-k)} - \frac{A^{(1)}}{r_+\be}\,,
\end{align}
and we used the fact that
\begin{equation}
    \De = \be^2\frac{f}{(1-f)^2} \,, \qquad r- r_- = \frac{\be}{1-f}\,.
\end{equation}
If we fix a single modification $k$, we can get rid of the rational behaviour in $f$ by multiplying the equation by $( 1 - \eta f )^k$, obtaining the following expression
\begin{equation}\label{eq:Leaver_Teuk_mod2}
\begin{split}
    \sum_{n = 0}^N & R_n \Bigg[ \left(\frac{\al^\note{bg}_{n-1}}{f} + \be^\note{bg}_{n} + \ga^\note{bg}_{n+1} f \right)\left( 1 - \eta f \right)^k \\
    & +  \frac{\al^{(-k)}}{\be^2} \frac{(1-f)^{k+2}}{f}  \Bigg] f^n = 0\,.
\end{split}
\end{equation}
We can figure out the coefficient relation (equivalent to that of equation~\eqref{eq:coeff_relation}), which at a given $n$ takes the form
\begin{equation}
    \sum_{j = -1}^{k+1} \left( \widetilde{\ga}_{n,j-1} + \widetilde{\be}_{n,j} + \widetilde{\al}_{n,j+1} \right) R_{n-j} = 0\,.
    \label{recurrence_relation}
\end{equation}
The coefficients appearing in the relation are given by
\begin{align}
    \widetilde{\al}_{n,j} & = \left(-\eta\right)^{j} \binom{k}{j} \al^\note{bg}_{n-j} + (-1)^j \binom{k+2}{j} \frac{\al^{(-k)}}{\be^2}\,, \\
    \widetilde{\be}_{n,j} & = (-\eta)^j \binom{k}{j} \be^\note{bg}_{n-j}\,, \\
    \widetilde{\ga}_{n,j} & = \left(-\eta\right)^{j} \binom{k}{j} \ga^\note{bg}_{n-j} \, .
\end{align}
We notice that, from the definition of the binomial, $\widetilde{\al}_{n,j}$ is non-vanishing for $0 \leq j \leq k+2$, while $\widetilde{\be}_{n,j}$ and $\widetilde{\ga}_{n,j}$ are non-zero for $0 \leq j \leq k$.
Now that we have a $k+3$ terms relation, we can perform a Gaussian elimination to reduce it to a three-terms relation (details can be found in the appendix of~\cite{Volkel:2022aca}). Once the three-terms relation is found, one can re-initialize the ladder operator $\La^r_n$ and obtain the modified frequency and separation constant from equations~\eqref{eq:Leaver_condition_r}--\eqref{eq:Leaver_condition_th}.

\subsection{Numerical computation of the coefficients}

In the previous two sections we explained how to obtain the functions $\mathcal{L}_r\left(\om, B, \al \right)$ and $\mathcal{L}_\th\left(\om, B \right)$. To compute the coefficients $d_\om$ and $d_B$ as given in equation~\eqref{eq:coefficients_d}, we evaluate the derivatives numerically with a 4-points centered stencil. For each pair of coefficients, we initialize the ladder operators $\La_N$ to some arbitrary low integer $N$, and then increase it by one until the simultaneous relative change in $d_\om$ and $d_B$ is smaller than a given tolerance, which we chose to be $10^{-7}$ \tcb{(the same accuracy has been required for the computation of the frequencies)}. 
We computed numerically all the coefficients for the following values $s=-2$, $n=[0,2]$, $\ell=[2,4]$, $m=[-\ell,\ell]$, $k=[-3,4]$ in a uniform grid in $a=[0,0.495]$ with spacing $\de a =0.005$. The full list of coefficients is available in a public git folder~\cite{github}.

\begin{figure*}
\centering
\includegraphics[width=0.75\linewidth]{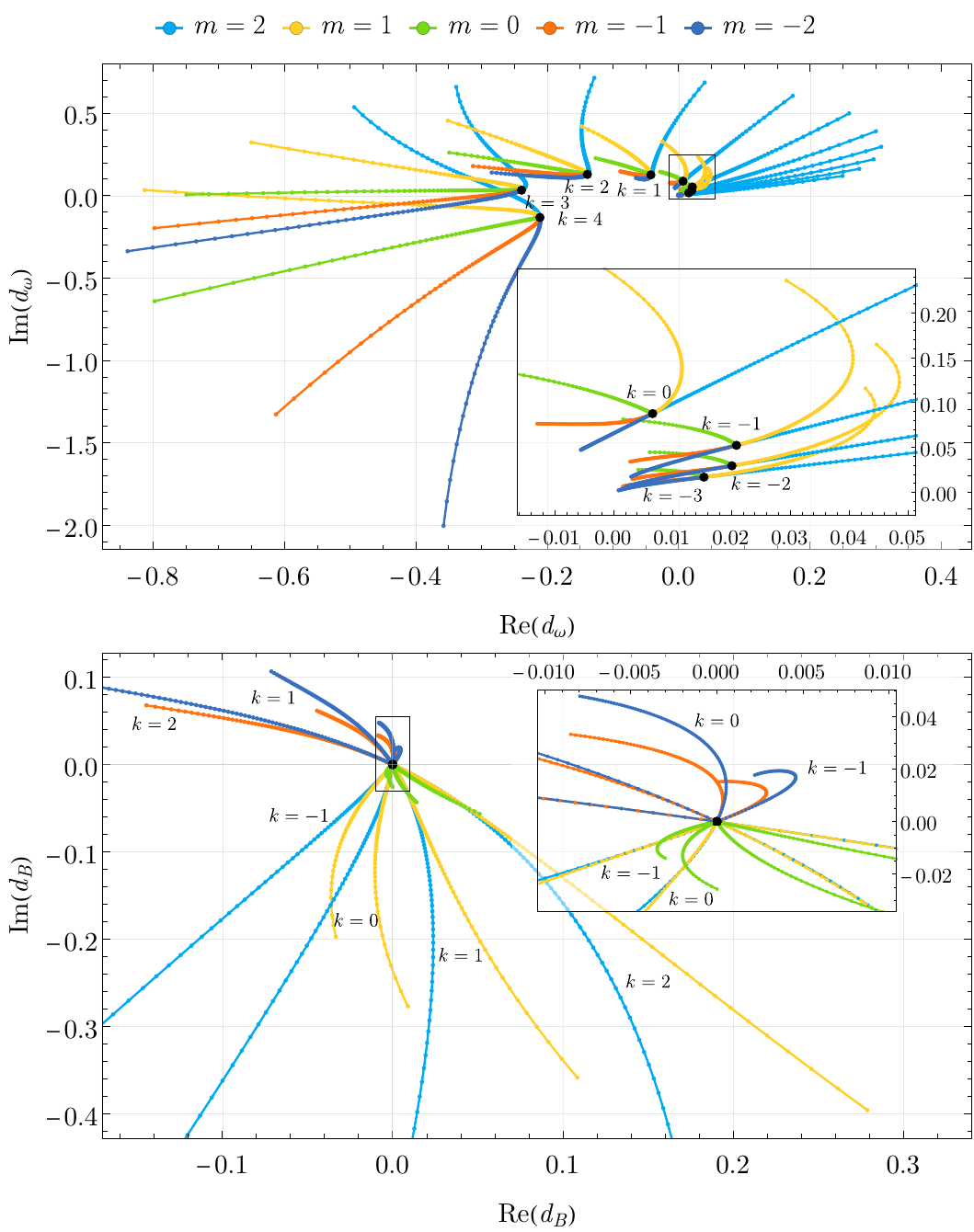}
\caption{In the top panel we show the real and the imaginary part of $d_\om$ for $n=0$, $\ell=2$, $m=[-2,2]$, $k=[-3,4]$ and values of the spin from $a=0$ to $a=0.45$, and each point is on a step of $\de a=0.005$. The inset focuses around the coefficients with $k=[-3,0]$. In the bottom panel we show the real and the imaginary part of $d_B$ for $n=0$, $\ell=2$, $m=[-2,2]$, $k=[-1,2]$ and same values of the spin. The inset focuses around the coefficients with $m\leq0$ and $k=[-1,0]$. In both plots the black dot signals the coefficient value for $a=0$. \label{Fig:domega}
}
\end{figure*}

In figure~\ref{Fig:domega} we show the results from this computation for the real and imaginary parts of the $d_\om^{(k)}$ coefficients for $s = -2$, $n=0$, $\ell = 2$, $m=[-2,2]$ for values of $k=[-3,4]$ and of the spin $a$ comprised between $0$ and $0.45$, as well as the real and the imaginary part of  $d_B^{(k)}$ for the same $n,\ell,m$ and $k=[-1,2]$.

To directly apply our formalism to further studies, e.g., ringdown analysis of non-linear computations or data analysis, we also provide a \texttt{python} code and a \texttt{jupyter notebook} with some examples~\cite{github}. It allows one to compute the QNMs and the separation constants as function of $n,\ell, m, a$ and $\alpha^{(k)}$ and can thus, in principle, be efficiently integrated in commonly used code infrastructure. 
The code also allows one to access some of the earlier results for the parametrized QNM framework for modifications to the Regge-Wheeler and Zerilli potentials, for which coefficients beyond the fundamental mode have been computed in reference~\cite{Volkel:2022aca}. The GR values for the QNMs have been taken from reference~\cite{Berti:2009kk,Berti:2005ys}. For more details about how the code is structured and how it can be used, we refer to the provided tutorial.

In principle, one should be able to compare the coefficients for $a=0$ with those computed in~\cite{Cardoso:2019mqo,Volkel:2022aca}. However, we stress that for $a=0$, equation~\eqref{eq:Teukolskymod} reduces to the non-spinning limit of the Bardeen-Press equation~\cite{Bardeen:1973xb}, whereas the formalism of~\cite{Cardoso:2019mqo,Volkel:2022aca} was developed for the Regge-Wheeler and the Zerilli equation. The transformation between the Bardeen-Press potential and the Regge-Wheeler/Zerilli potentials was obtained by Chandrasekhar~\cite{Chandrasekhar:1975nkd}, but generalizing this to the case of the modified potential with generic $\al^{(k)}$ couplings is non-trivial. 

\subsection{Linearized regime of validity}
\label{Linearized_regime}
The framework we developed is motivated by the assumption that any modification of gravity produces only slight deviations from GR in astrophysical observables. In this section, we expand on the regime of validity of the formalism, by providing a quantitative assessment of the accuracy of such approach. It is worth noting that we can only asses the error made by restricting to linear corrections to the frequencies, as defined in equation~\eqref{eq:dOmega_dB} and not taking into account higher-order corrections to the potential, which are beyond the scope of this paper. 

\begin{figure*}
\centering
\includegraphics[width=\textwidth]{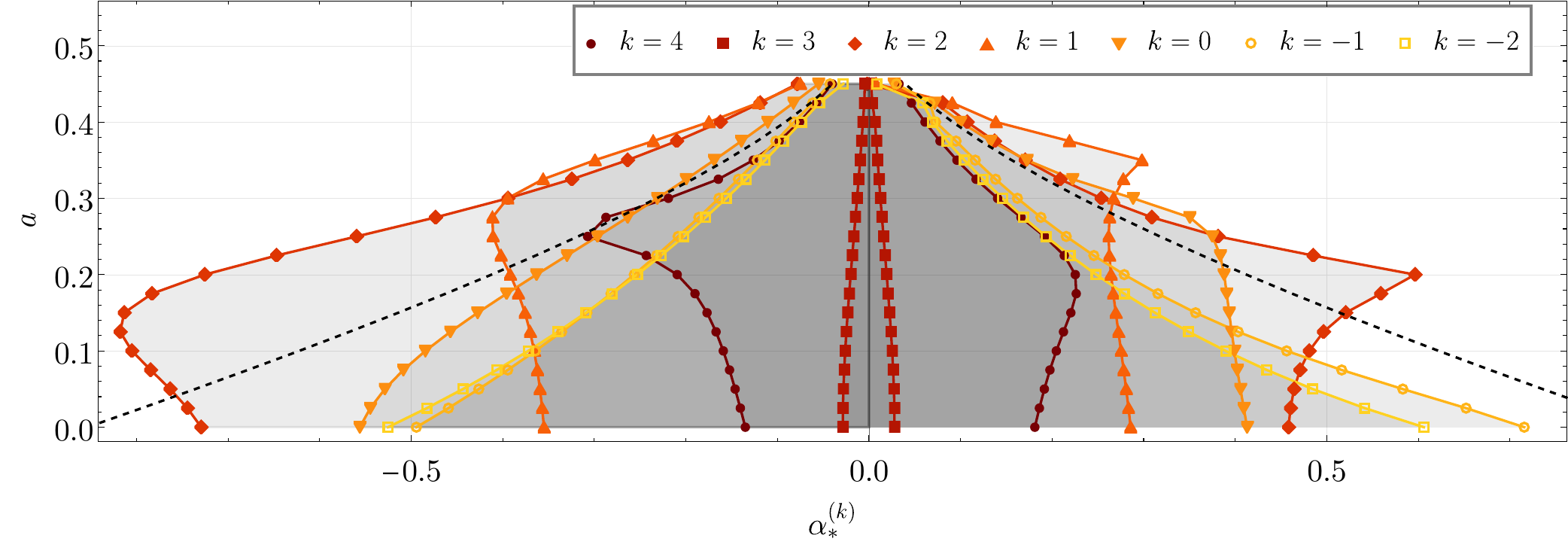}
\caption{We plot the threshold values $\alpha^{(k)}_*$ against the spin for different values of $k$. With $\alpha^{(k)}_*$, we identify the limit value of a coupling at which an error of $1\%$ the a linear and a non-linear approximation is obtained. The plot uses as non-linear estimates the full continued fraction results. The error is evaluated for the $n=0$, $\ell=m=2$ mode and different values of $k$ and of the spin $a$. The black dashed lines correspond to the estimate~\eqref{eq:estimate}}
\label{thresholds}
\end{figure*}

First of all, we give a heuristic motivation on the maximum size of the coefficients, by requesting that the perturbation equation is not strongly modified at the boundaries of our dominion and that $\om\simeq \Ord(1)$. At $r\to\infty$, we have seen from equations~\eqref{eq:infty_q} and~\eqref{eq:infty_p} that the only coefficients modifying the asymptotic structure of the potential are $\al^{(3)}$ and $\al^{(4)}$. With some simple algebra, we can infer
\begin{equation}
    \al^{(3)} \lesssim 1 \,, \qquad \al^{(4)}  \lesssim 1 \,.
\end{equation}
On the other hand, the modifications in the potential affect the near-horizon expansion as in equation~\eqref{eq:hor_expansion}. The condition is such that the sum of coefficients must behave as
\begin{equation}\label{eq:estimate}
    \sum_k\al^{(k)} \lesssim \left| 2\be^2 \sg_\note{GR} \left( \sg_\note{GR} - \frac{\ii s}{2} \right) \right| \,.
\end{equation}
It is worth noting that when the superradiant condition $\om=\om_c$ is activated, one has $\sg_\note{GR} = 0$, and we expect that the formalism is valid only if the sum of the $\al^{(k)}$ is approximately 0.

In general, however, each power of $k$ affects in a different way the effective potential. In order to have a more quantitative estimate of the allowed regime of validity, we perform two separate analysis. 
First, we compare the QNM frequencies computed with the linear approximation against those obtained with a full continued-fraction method discussed above.
We estimated the error on the frequencies as
\begin{equation}
    \De_\omega \equiv \sqrt{\left(\frac{\De \omega_R}{\omega_R}\right)^2+\left(\frac{\De \omega_I}{\omega_I}\right)^2}\,,
\end{equation}
where $\omega_{R,I}$ are respectively the real and imaginary parts of the QNM. We computed the error for several real positive and negative values of the couplings $\alpha^{(k)}$ and extracted the threshold values $\alpha^{(k)}_*$ at which the error reaches $1\%$.
The results are represented in figure~\ref{thresholds} for the mode $n=0$, $\ell=m =2$, for selected values of the spins between $a=0$ and $a=0.45$ and for $k=[-2,4]$. 

It can be seen that there is a complicated dependence of the thresholds on the type of modification we introduce in the potential. However, a main qualitative feature can be read off, i.e. that, for any modification, the threshold tends to get smaller for higher spins. The physical interpretation of that, is that, for a given beyond-GR effect in the modified Teukolsky equation, rotation tends to exacerbate the deviation of the linear approximation with respect to the true values of the QNMs. Bearing this caveat in mind, we will still show in the next section that the linear approximation provides very good results in a couple of known models of perturbation of rotating BHs with deviations from Kerr, also for high spin.

Since the computation of QNMs with the continued fraction method is not immediate nor straightforward to implement, we want to provide a quick estimate for the errors of the single-$k$ contributions. In this respect, we compute the diagonal quadratic corrections, as explained in appendix~\ref{app:quadratic}. We checked the estimate $\al_*^{(k)}$ by computing the error $\De_\om$ assuming that the non-linear frequencies are obtained including quadratic coefficients. By a qualitative comparison, the quadratic estimate works well to capture the error except for $k=3$, and partially for $k=4$. Even though it is not as precise as the full non-linear comparison, the quadratic coefficients can be used as a quick way to understand what threshold value to take for the couplings.

Lastly, we want to stress that the thresholds that we provided in this section, are referred to the contribution of a single modification. Hence, it could be that, depending on the values of the coefficients, the combination of multiple $k$ would need larger or smaller threshold values. This means that for a theory-specific case, the bounds on $\al^{(k)}$ might differ from what we inferred in this section, and need to be addressed case-by-case.

\section{Applications}\label{sec:applications}

\subsection{Massive scalar perturbations}
\begin{figure}
\centering
\includegraphics[width=\linewidth]{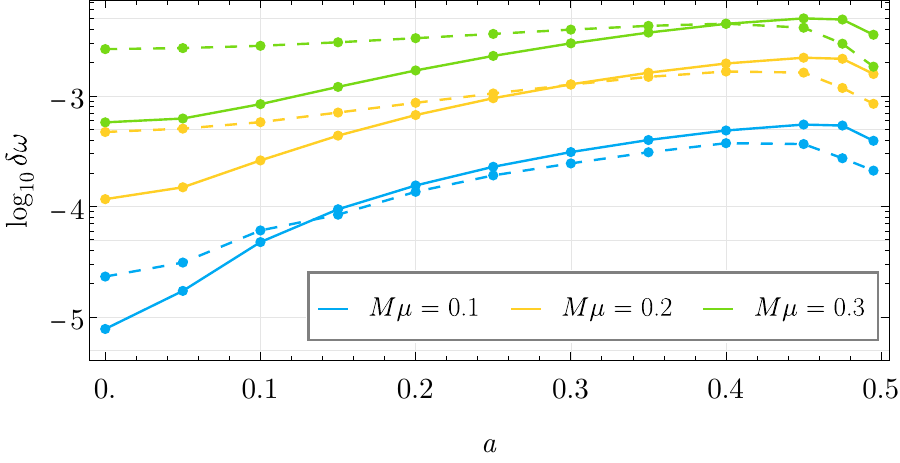}
\caption{ Relative difference for the real part (solid line) and imaginary part (dashed line) of the fundamental $\ell=m=2$ mode for a massive scalar perturbation computed with the linear approximation against the non-linear results of~\cite{Dolan:2007mj}. \label{Fig:massive}
}
\end{figure}

The first example that we provide to test our formalism is for the computation of the QNMs of a massive scalar field, a case  extensively studied in the literature~\cite{Zouros:1979iw,Detweiler:1980uk,Dolan:2007mj}.
The radial and angular perturbation equations for a massive scalar field ($s=0$) with mass $\mu$ are
\begin{align}
    \frac{\dd}{\dd r} & \left[ \De R'(r) \right] + \left( \frac{K^2}{\De} - \la_{\ell m} -\mu^2 r^2 \right) R(r) = 0\,, \\
    \frac{\dd}{\dd y} & \left[ \left(1-y^2\right) S'(y) \right] \notag \\
    & + \left[a^2\left(\om^2 - \mu^2 \right) y^2 + B_{\ell m} - \frac{m^2}{1-y^2}\right]S(y) = 0\,.
\end{align}
First of all, we bring the angular equation into the form of equation~\eqref{eq:Teuk_ang} by transforming $\om \to \om + \frac{\mu^2}{2\om}$. Then, by assuming $\mu \ll 1$, the radial equation is automatically brought in the form of~\eqref{eq:Teukolskymod}, with the only non-zero $\al^{(k)}$ being
\begin{equation}
    \al^{(1)} = \mu^2 a r_+ \left( a - \frac{m}{\om_0} \right)\,, \qquad \al^{(3)} = \mu^2 r_+^3\,,
\end{equation}
where $\om_0$ is the unperturbed Kerr frequency. The effect of the mass on the frequency at linear order in $\mu^2$ is given by
\begin{equation}\label{eq:om_lin_massive}
    \om_\note{L} = \om_0 + \frac{\mu^2}{2\om_0} + d_{(1)} \al^{(1)} + d_{(3)} \al^{(3)}\,.
\end{equation}
In figure~\ref{Fig:massive} we show the difference $\de\om = |\om_\note{L} - \om_\note{NL}|$ between the linear results in~\eqref{eq:om_lin_massive} and the nonlinear QNMs $\om_\note{NL}$ computed in~\cite{Dolan:2007mj} for $\ell=m=2$ modes.

\subsection{The Dudley-Finley equation}

\begin{figure*}
\centering
\includegraphics[width=\linewidth]{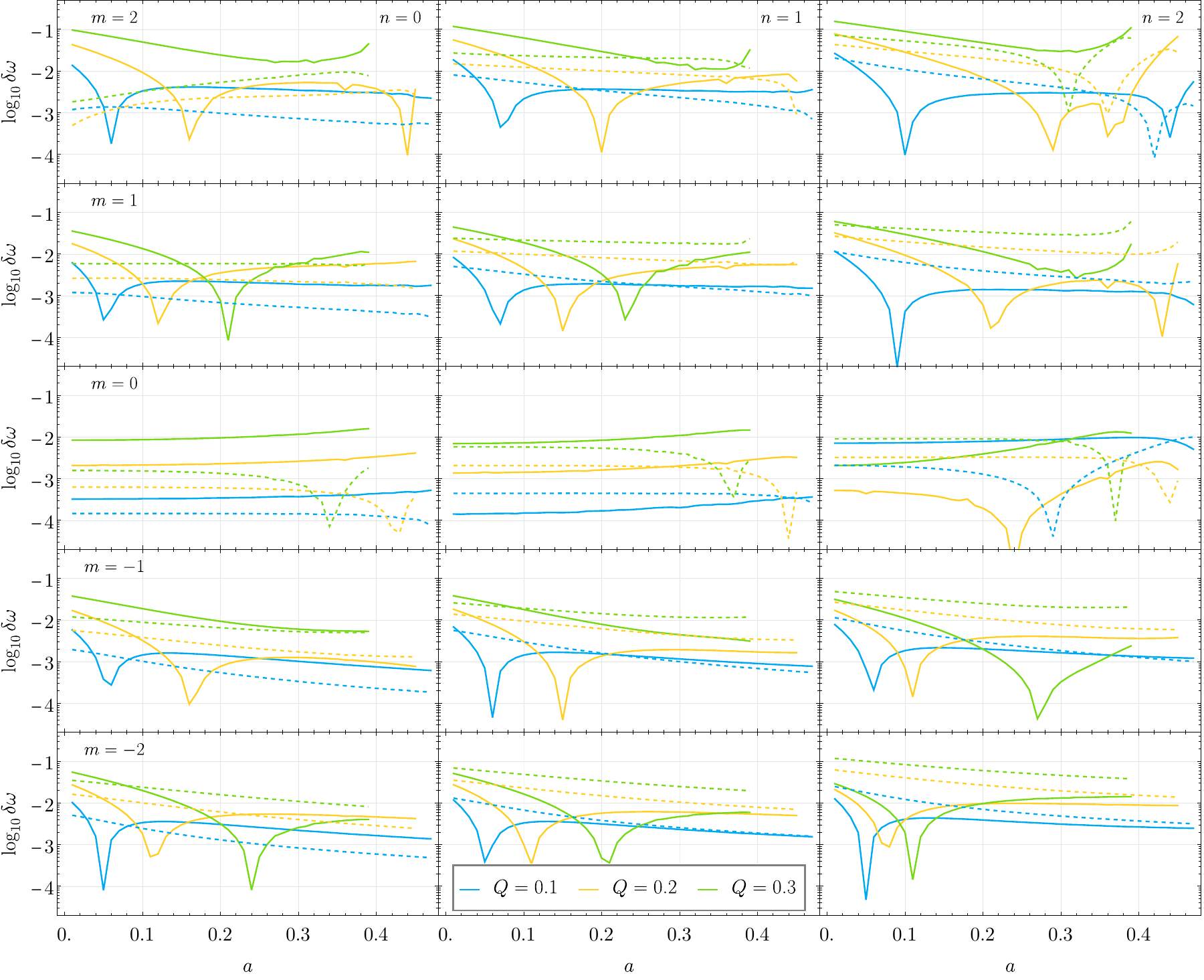}
\caption{Difference between the real (solid lines) and imaginary (dashed lines) part of the Dudley-Finley QNMs computed either with the linear perturbative approach or the full continued fraction method. We show results for $n=0$ (left panels), $n=1$ (central panels) and $n=2$ (right panels), $\ell = 2$, $m=[-2,2]$ (bottom to top panels) for different values of the spin and of the electric charge. Note that the three curves have different endpoints, as for a given $Q$ the maximum value of $a$ is $a_\note{max} = \frac{1}{2}\sqrt{1-4Q^2} $}
\label{Fig:DF_comparison}
\end{figure*}

As a second example, we would like to test our formalism against gravitational perturbation of a Kerr-Newman (KN) BH in the limit of small charge, since the QNMs for a generic electric charge $Q$ have been computed numerically in~\cite{Dias:2021yju} and fits are available in~\cite{Carullo:2021oxn}. Unfortunately, the KN perturbation equation is not explicitly separable, not even in the limit of small charge~\cite{Chandrasekhar:1984siy} in which at least electromagnetic and gravitational perturbations decouple. One could apply the algorithm of~\cite{Li:2022pcy,Hussain:2022ins,Cano:2023tmv} to obtain a modified Teukolsky operator for the KN solution, but it goes beyond the scopes of this paper. For the sake of testing the method, we can restrict ourselves to the Dudley-Finley (DF) equation, a proxy equation for the perturbations of the Kerr-Newman metric~\cite{Dudley:1977zz,Dudley:1978vd,Berti:2005eb}. The DF equation is obtained by taking equation~\eqref{eq:Teukolsky} and performing the following substitution
\begin{equation}
    \De \to \De + Q^2 \,.
\end{equation}
We can then rescale the equation into the form~\eqref{eq:Teukolskymod} by assuming that $Q \ll 1/2$ and by defining a new spin parameter
\begin{equation}\label{eq:spin_rescaling}
    \widebar{a} = a + \frac{Q^2}{2a}\,,
\end{equation}
such that $\widebar{a}^2 \simeq a^2 + Q^2$. Retaining only the terms quadratic in $Q$, we can see that the only non-zero $\al^{(k)}$ terms that contribute to the equation~\eqref{eq:delta_potential} are
\begin{align}
    \al^{(0)} & = Q^2 \left[ \frac{\ii s}{2 \widebar{a}}(m - 2 \widebar{a} \widebar{\om}_0) - (m - \widebar{a} \widebar{\om}_0)^2 \right] \,, \\
    \al^{(1)} & =  Q^2 r_+ \left[\widebar{\om}_0\frac{m - \widebar{a} \om_0}{\widebar{a}} -\frac{\ii s}{\widebar{a}}(m - 2 \widebar{a} \widebar{\om}_0) \right]\,, \\
    \al^{(2)} & = -Q^2 r_+^2 \widebar{\om}_0^2 \,,
\end{align}
where $\widebar{\om}_0$ is the Kerr frequency evaluated at spin $\widebar{a}$. The DF linear frequencies at spin $a$ are obtained by
\begin{equation}\label{eq:DF_freq}
    \om_\note{L} = \widebar{\om}_0 + \sum_{k=0}^2 \al^{(k)} d_{(k)} \,.
\end{equation}
In Figure~\ref{Fig:DF_comparison} we show the real and imaginary part of the absolute difference $\de\om = |\om_\note{L} - \om_\note{NL}|$ between the linear results in~\eqref{eq:DF_freq} and the nonlinear QNMs $\om_\note{NL}$ computed via the Leaver method in~\cite{Berti:2005eb}, for various spins and different values of the electric charge. We make the comparison with $\ell=2$ modes, with all values of $m=[-2,2]$, for the fundamental and first two overtones. The plot clearly shows that the discrepancy between the linearized QNMs and the full non-linear results scales with the charge, and the approximation remains valid for all the different values of $(n,\ell,m)$ surveyed. 

Finally, we comment on the fact that the errors grow for small values of the spin. This is due to the fact that in order to bring the equation in the form of~\eqref{eq:Teukolskymod}, we performed the transformation~\eqref{eq:spin_rescaling}, which brings a term $1/a$ to the denominator when $m\neq 0$. In other words, this transformation is valid as long as $|Q|\ll |a|$. Nevertheless, the smallness of the universal coefficients $d_{(k)}$ is such that the combination in frequency~\eqref{eq:DF_freq} is finite and faithful to the non-linear value.

\subsection{Higher derivative gravity}

\begin{figure}
\centering
\includegraphics[width=\linewidth]{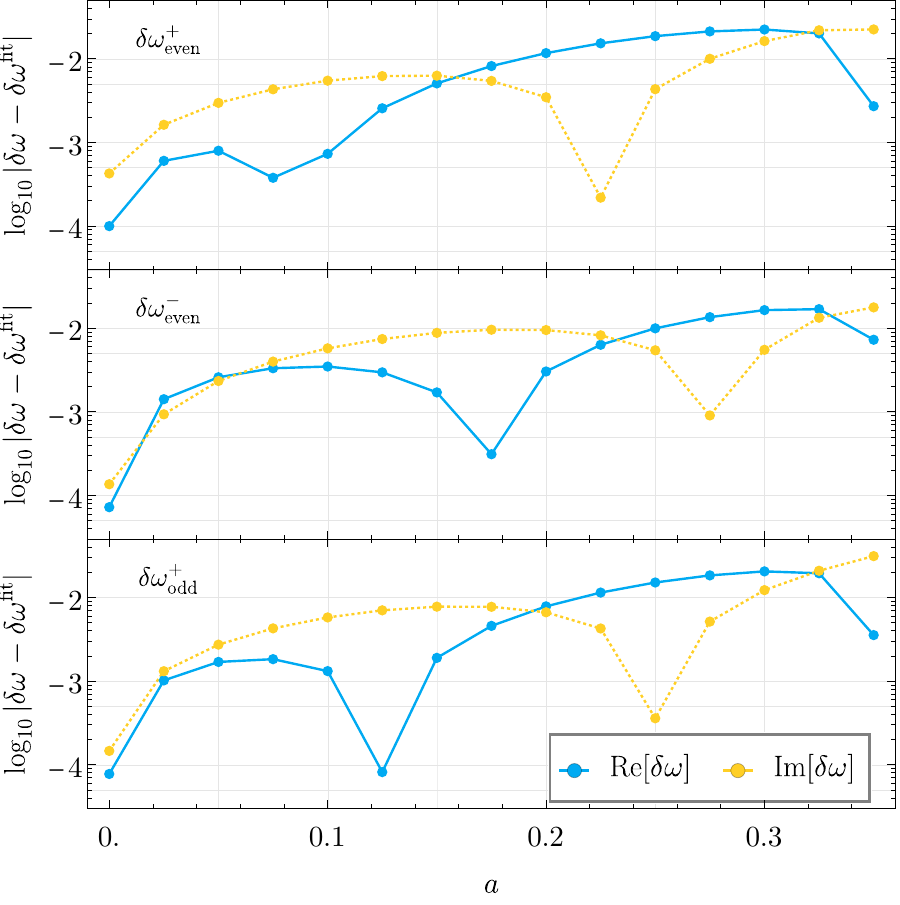}
\caption{Real and imaginary part of the absolute difference between the fits of~\cite{Cano:2023jbk} and the frequencies computed with the parametrized formalism for the $\pm$ polarizations of even cubic modes and the $+$ polarization of the odd cubic mode
\label{Fig:HD_pm}
}
\end{figure}

Now we want to check the prediction of QNMs in higher derivative gravity using the parametrized method against the results presented in~\cite{Cano:2023jbk}. In a companion paper~\cite{Cano:2024ezp}, focused on the analysis of QNMs in higher-derivative gravity, we show how to reduce the radial perturbation equation to the form of equation~\eqref{eq:Teukolskymod}, with the only non-vanishing values of $\al^{(k)}$ being $k=k^\note{HD}=[-2,0,1,2]$
\begin{equation}
   \de V^\pm = \la \sum_{k\in k^\note{HD}} \al_\pm^{(k)} \left( \frac{r}{r_+}\right)^k\,,
\end{equation}
where the $\pm$ refers to the polarization of the perturbation and we collected out $\la$, the normalized coupling constant of the theory.\footnote{\textit{cfr.}~equation~(30) of~\cite{Cano:2024ezp}, for which $\al^{(k)} = A^{(k)} r_+^k$, and the coupling constant has been previously factorized out. Here we use $\la$ to refer to the coupling constant, to avoid misunderstanding with the $\al_\mathrm{q}$ used in~\cite{Cano:2024ezp}.}
From this, we can compute the frequencies deviations, normalized by the coupling constant $\la$
\begin{equation}
    \de \om^{\pm} = \frac{\om^\pm - \om^\note{Kerr}}{\la} = \sum_{k \in k^\note{HD}} \al^{(k)}_\pm d_{(k)}\,,
\end{equation}
for each parity, and each realization of the theory. In figure~\ref{Fig:HD_pm}, we compare our results against the fits $\de\om^\mathrm{fit}$ given in~\cite{Cano:2023jbk}. We truncate the analysis at spin $a=0.35$, since the fits are valid only up to this value. The plot shows remarkable agreement between the corrections computed with two different methods, strengthening the validity of the parametrized formalism. In figure~\ref{Fig:HD_pm} we limited to show polar and axial, $\ell=m=2$ values for the even-parity cubic theory, labelled as $\de\om_\mathrm{even}^\pm$, as well as polar, $\ell=m=2$ values for the odd-parity cubic theory, labelled as $\de\om_\mathrm{odd}^+$. Details on the definition of these modes can be found in~\cite{Cano:2023jbk} and in the companion paper where we perform an extensive study of QNMs of rotating BHs in higher derivative gravity~\cite{Cano:2024ezp}.

\section{Conclusions}\label{sec:conlusions}

In this paper we have shown how to connect small deviations parametrized by powers of the radial coordinate $r$ in the Teukolsky equation to small deviations in the eigenfrequencies and in the separation constants of modified Kerr BHs. We proved that for each value of $n,\ell,m$ there are up to nine independent coefficients in the radial parametrization, but for specific cases they could be less. We presented a robust method to compute the coefficients that control the linear corrections to the QNM frequencies and separation constants through a generalization of Leaver's continued fractions, and used it to compute them for $n=[0,2]$, $\ell=[2,4]$, $m=[-\ell,\ell]$ and $k=[-3,4]$ in a range of spins between $0$ and $0.495$.\footnote{We recall that in our conventions extremality corresponds to $a=1/2$} These results are available online in a public git repository~\cite{github}, together with a \texttt{python} code and a \texttt{jupyter notebook} to compute the QNMs and separation constants. There we also provide a tutorial demonstrating how to use the code, which can in principle be applied to compute QNMs with arbitrary $n$, $l$, $m$, $k$ and angular momentum besides those we computed explicitly here. 

We checked the quality of the predictions against three cases known in the literature: perturbations of a massive scalar field around Kerr, the Dudley-Finley equation and the QNMs of BHs in higher-derivative gravity. For all the three cases the frequencies predicted by the formalism show great agreement with respect to the results in the literature.

These coefficients will be particularly useful for the computation of QNMs of rotating BHs in alternative theories of gravity for which a modified Teukolsky equation is obtained, in the spirit of the method developed in~\cite{Li:2022pcy,Hussain:2022ins,Cano:2023tmv}. 
So far, this method has been successfully applied to higher-derivative gravity \cite{Cano:2023jbk}, but other theories like scalar-Gauss-Bonnet gravity and dynamical-Chern-Simons gravity \cite{Wagle:2023fwl} are good candidates for this computation. In order to study those cases, it would be interesting to generalize our parametrized formalism by including couplings between the Teukolsky equation and a scalar field, analogous to the analysis of \cite{McManus:2019ulj} in the case of static BHs. 

The modified Teukolsky approach of \cite{Li:2022pcy,Hussain:2022ins,Cano:2023tmv} is currently limited by the fact that metric reconstruction is only available for GR, meaning that it is not yet possible to extend the method beyond first order in the coupling. For this reason, we did not explore further the quadratic coefficients as it was done in the non-rotating case~\cite{McManus:2019ulj,Volkel:2022aca}, and we limited the computation of the diagonal ones just to have a quick estimate of the error of the method itself.

Let us also remark that a different approach to study beyond-GR QNMs based on spectral methods has recently been introduced and successfully demonstrated for a wide range of spins in Refs.~\cite{Chung:2023zdq,Chung:2023wkd,Blazquez-Salcedo:2023hwg,Chung:2024ira,Chung:2024vaf}. 
These methods have the advantage of being more flexible, but on the other hand, they have a much higher computational complexity and cost than the standard perturbative approaches. In this regard, the results of our method applied to specific theories may be useful to validate the spectral approaches.

The most intriguing open problem from our analysis is whether one can find a better way to exploit the potential ambiguity, as done in section~\ref{sec:ambiguity}. With the choice we made, we could reduce the number of independent coefficients $\al^{(k)}$ to 8. For the case of higher derivative gravity we have been able to numerically reduce the number of free coefficients to just 4~\cite{Cano:2024ezp}. One may wonder whether higher derivative gravity has a special structure of the equations, or if there is a fundamental transformation of the potential that could diminish the number of free parameters.

Such discussion is relevant especially if one wants to use this formalism in a theory-agnostic setup, \eg, to perform ringdown tests of GR or the inverse problem. Since the coefficients $d^{(k)}$ have a spin dependence, it would be interesting to map them to ParSpec~\cite{Maselli:2019mjd}. Another useful mapping would be with the WKB deviation coefficients, as done in~\cite{Volkel:2022khh}. Finally, in the upcoming analysis, it would be interesting to compare the detectability of beyond-Teukolsky effects against that of second order QNMs~\cite{Mitman:2022qdl,Cheung:2022rbm,Yi:2024elj}.

\acknowledgments

We thank Emanuele Berti for fruitful discussions. We also thank Adrian Ka-Wai Chung, Pratik Wagle and Nicolas Yunes for useful comments.  The work of P.~A.~C.~received the support of a fellowship from “la Caixa” Foundation (ID 100010434) with code LCF/BQ/PI23/11970032. S.~M.~acknowledges support from the Flemish inter-university project IBOF/21/084. S.~H.~V.~acknowledges funding from the Deutsche Forschungsgemeinschaft (DFG): Project No. 386119226. The work of L.~C.~was supported by the European Union’s H2020 ERC Consolidator Grant ``GRavity from Astrophysical to Microscopic Scales'' (Grant No. GRAMS-815673), the PRIN 2022 grant ``GUVIRP - Gravity tests in the UltraViolet and InfraRed with Pulsar timing'', and the EU Horizon 2020 Research and Innovation Programme under the Marie Sklodowska-Curie Grant Agreement No. 101007855.

\appendix

\section{Ambiguity of the potential modifications}\label{app:ambiguity}
Here we list the explicit form of the coefficients of equation~\eqref{eq:ambiguous} when one transforms the radial Teukolsky function as~\eqref{eq:transformation_ambiguous}. These values hold for $s=-2$:
\begin{widetext}
\begin{align}
\overline{A}_j^{(-3)} = & \, \frac{a^6}{2} j (j+1)(j+2) \,, \qquad
\overline{A}_j^{(-2)} = \, -\frac{3a^4}{2} j (j+1)^2 \,, \\
\overline{A}_j^{(-1)} = & \, \frac{a^4}{2}  j \left[3 j (j+1)-4 B +4 m^2-8 i \om +10\right] +4 i a^3 j m+\frac{3}{2} a^2 j \left(j^2+j-1\right) \,, \\
\overline{A}_j^{(0)} = & \, a^3 (2 j-1) \left[a \om  (\om -4 i)-2  m (\om +2 i) \right] \notag \\
& - a^2 \left[3 j^3 - j \left(4 B -2 m^2+4 i \om -1\right) + B + 2\right] -4 i a j m-\frac{1}{2} j
   \left(j^2-4\right) \,, \\
\overline{A}_j^{(1)} = & \, 2 a^4 (j-1) \om ^2 +\frac{1}{2} a^2 \Big[3 j^2(j-1)+4 B +2 \om  (\om -16 i)+8 +j \left(-8 B +4 m^2-4 \om  (\om -8 i)+2\right) \Big] \notag \\
& +2 a m \left[2 j
   (\om +3 i)-\om -2 i\right] +\frac{1}{2} j \left[3 (j-1) j-4 B -11\right]+B +2 \,, \\
\overline{A}_j^{(2)} = & \, a^2 \om  \left[\om -16 i (j-1)\right]-2 a (2 j-1) m (\om +2 i) +\frac{1}{2} j (-3 (j-2) j+8 B -24 i \om +13)-3 (B -4 i \om +2) \,, \\
\overline{A}_j^{(3)} = & \, 2 a^2 (2 j-3) \om ^2+\frac{1}{2} (j-1) \left[(j-2) j-4 (B +2)\right]+4 i (5 j-6) \om \,, \\
\overline{A}_j^{(4)} = & \, 2 \om \left[2 (\om +3 i)- j   (\om +4 i) \right] \,, \qquad
\overline{A}_j^{(5)} =  \,  2 (j-2) \om ^2 \,.
\end{align}    
\end{widetext}

\section{Nollert's improvements of continued fraction}\label{app:Nollert}
The procedure to numerically solve Teukolsky equation through Leaver's method requires in practice an initialization for the radial ladder operator $\Lambda^r_n$. Such quantity can be expanded for large initialization number $N$ as
\begin{equation}
    \Lambda^r_N= \sum_{j=0}^J C_j N^{-j/2} + \mathcal{O}\left(N\right)^{-\left(J+1\right)/2} \,.
    \label{large_n_exp}
\end{equation}
One can initialize the ladder operator just retaining the first term $C_0=-1$. However, this approximation requires in general a very high initial value for $N$ (which means long computational time) and appears to be insufficient for frequencies with large imaginary part (namely higher overtones). In~\cite{Nollert:1993zz}, it was shown that adding further corrections to the initial $\Lambda^r_N$ improves the accuracy of the method, also allowing to capture higher overtones. The $(k+3)$-terms recurrence relation of equation~\eqref{recurrence_relation} for $n=N$ can be expressed as
\begin{equation}
    \sum_{j=-1}^{k+1}M_{N,j}\,R_{N-j} = 0 \,,
\end{equation}
where we defined
\begin{equation}
    M_{N,j}\equiv  \widetilde{\ga}_{N,j-1} + \widetilde{\be}_{N,j} + \widetilde{\al}_{N,j+1}\,.
\end{equation}
Dividing it by $R_{N-k-1}$ and using the definition of ladder operator $\Lambda^r_N=-a_{N+1}/a_N$ one obtains the equation
\begin{equation}
    \sum^{k+1}_{j=-1}(-1)^j\,M_{N,j}\,\prod_{i=j+1}^{k+2}\Lambda^r_{N-i}=0\,.
\end{equation}
Plugging the definition~\eqref{large_n_exp} into the above formula, one can solve for the coefficients order by order. By the definition of $ \widetilde{\ga}_{N,j}$, $\widetilde{\be}_{N,j}$ and $ \widetilde{\al}_{N,j}$, they scale with $N$ as
\begin{align}
    \widetilde{\al}_{N,j} & = N^2 + \widetilde{\al}_1 N + \widetilde{\al}_0 \\
    \widetilde{\be}_{N,j} & = -2N^2 + \widetilde{\be}_1 N + \widetilde{\be}_0 \\
    \widetilde{\ga}_{N,j} & = N^2 + \widetilde{\ga}_1 N + \widetilde{\ga}_0
\end{align}
If we fix $C_0=-1$, then the other coefficients up to $J=5$ can be written as
\begin{align}
    C_1 = & \, \pm\sqrt{- \widetilde{\al}_1 - \widetilde{\be}_1 - \widetilde{\ga}_1} \,, \\
    C_2 = & \, \widetilde{\al}_1 + \frac{\widetilde{\be}_1}{2} - \frac{1}{4}  \,, \\
    C_3 = & \, \frac{C_2^2}{2 C_1} - \frac{C_2}{4 C_1} -\frac{ \widetilde{\al}_0 +\widetilde{\be}_0+\widetilde{\ga}_0}{2 C_1} - \frac{1 + 2 \widetilde{\al}_1}{4} C_1 \,, \\
    C_4 = & \, \widetilde{\al}_0 - \frac{4 \widetilde{\al}_1 - 8 \widetilde{\be}_0 + 1}{16} - \frac{1+4\widetilde{\al}_1}{4} C_2-\frac{C_3}{2 C_1} \,, \\
    C_5 = & \, \frac{2 \widetilde{\al}_1 +3}{4 C_1}C_2^2 - \frac{8 \widetilde{\al}_0 + 4 \widetilde{\al}_1 +3}{16} C_1 -\frac{C_3^2}{2 C_1} \notag \\
    & -\frac{4 \widetilde{\al}_1 +3 }{16 C_1}C_2 + C_3
   \left(\frac{C_2}{2 C_1^2}-\widetilde{\al}_1-1\right) \notag \\
   & +\left(\frac{C_2}{C_1}-\frac{3}{4 C_1}\right) C_4 \,,
\end{align}
and the sign of $C_1$ is chosen such that $\mathrm{Re}\left(C_1\right)>0$.

We can do the same expansion for the angular part, by expanding
\begin{equation}
    \Lambda^\th_N= \sum_{j=0}^J D_j N^{-j/2} + \mathcal{O}\left(N\right)^{-\left(J+1\right)/2} \,.
\end{equation}
From equations~\eqref{eq:coeff_th_1}--\eqref{eq:coeff_th_3} we can schematically say that
\begin{align}
    \al^\th_N & = -2N^2 + \al_1 N + \al_0 \\
    \be^\th_N & =   N^2 + \be_1 N + \be_0 \\
    \ga^\th_N & =         \ga_1 N + \ga_0
\end{align}
By solving perturbatively in $1/N$ the relation~\eqref{eq:cf_relation}, we obtain the following expression for the coefficients $D_j$ up to $J=4$
\begin{align}
    D_0 = & \, 0 \\
    D_1 = & \, \ga_1 \\
    D_2 = & \, \ga_0 - \ga_1\left( 1 + \be_1 + 2\ga_1 \right) \\
    D_3 = & \, \ga_1^2(\al_1-2) - \ga_1 (1+\be_0 +\be_1) \notag \\
    & - D_2(\be_1 + 4\ga_1 +2) \\
    D_4 = & \, \al _0 \ga _1^2+\frac{2 D_2^2 \left(\be _1+3 \ga _1+2\right)}{\ga _1} \notag \\
    & +D_2 \left(\be _0+\be _1+2 \ga _1+1\right) \notag \\
    & +D_3 \left(\frac{2 D_2}{\ga _1}-\be _1-4 \ga
   _1-2\right)
\end{align}

\tcb{
\section{Derivation of the boundary conditions}
\label{BCs}
The correct implementation of the continued fraction method requires that the ansatz that one assumes properly encodes the behavior of the solution close to the singular points of the equation. We will now show how such ans\"atze are properly derived in the beyond-Teukolsky case.}

\tcb{Following the discussion in \cite{Teukolsky:1973ha}, it is convenient to introduce the new master field
\begin{equation}
    Y(r)=\Delta^{\frac{s}{2}}\left(r^2+a^2\right)^{\frac{1}{2}}R(r)\,,
\end{equation} 
and the tortoise coordinate $r_*$ defined by
\begin{equation}
    \frac{{\rm d}r_*}{{\rm d}r}=\frac{r^2+a^2}{\Delta}\,,
\end{equation}
Note that this tortoise coordinate has the asymptotic behavior
\begin{equation}
\begin{split}
    &r_* \xrightarrow{r\to \infty}r\,,\\
    &r_*\xrightarrow{r\to r_+}\frac{r_+}{\beta}\ln(r-r_+)\,.
\end{split}
\end{equation}
This choice allows to rewrite the modified Teukolsky equation in a form in which the first derivative of the master variable does not appear.
For $r \to \infty $ it reads
\begin{equation}
\begin{split}
    \frac{{\rm d}^2}{{\rm d}r_*^2}Y+&\left[\left(\omega^2+\frac{\Tilde{\alpha}^{(2)}}{r_+^4}\right)\right.+\\
    &+\left.\left(2is\omega +\frac{\Tilde{\alpha}^{(1)}}{r_+^3}-\frac{\Tilde{\alpha}^{(2)}}{r_+^4}\right)r^{-1}\right]Y =0\,.
\end{split}
\end{equation}
According with the boundary conditions for quasi-normal modes, we choose the solution that is purely outgoing, which reads
$Y\sim r^{p-s}e^{qr_*}$, where the parameter $p$ and $q$ are the ones introduced in Eq. (\ref{eq:infty_q}) and (\ref{eq:infty_p}).}

\tcb{On the other hand, as $r\to r_+$, the modified Teukolsky equation reads
\begin{equation}
    \begin{split}
    \frac{{\rm d}^2}{{\rm d}r_*^2}Y+r_+^{-2}&\left[\left(K_+-i s\frac{\beta}{2}\right)^2+\sum_{k=-K}^4 \alpha^{(k)}\right]Y =0\,,
\end{split}
\end{equation}
where $K_+=r_+\omega-a m =\beta \sigma_{{\rm GR}}$. In this case, we want the solution to be purely ingoing, so we have
\begin{equation}
\begin{split}
    Y&\sim \exp{\left(-\frac{i r_*}{r_+}\sqrt{\left(K_+-i s\frac{\beta}{2}\right)^2+\sum_{k=-K}^4 \alpha^{(k)}}\right)}\\
    &\sim (r-r_+)^{-i \sigma-\frac{s}{2}}\,,
\end{split}
\end{equation}
where $\sigma$ is defined in Eq. (\ref{eq:hor_expansion}).
Considering the factor relating the master fields $Y$ and $R$, we obtain
\begin{equation}
    \begin{split}
        &R \xrightarrow{r\to \infty}e^{i q r}r^{p-2s-1}\,, \\
    &R\xrightarrow{r\to r_+}(r-r_+)^{-i\sigma-s}\,.
    \end{split}
\end{equation}
One can then easily verify that the ansatz of Eq. (\ref{ansatz}), with the generalized definitions of $\sigma$, $p$ and $q$, correctly encodes the behavior of the solution $R(r)$ at the boundaries. }

\section{Splitting of the potential}\label{app:potential}

In this section of the appendix we show how to transform the potential~\eqref{eq:delta_potential} into the potential~\eqref{eq:delta_potential_2}. We start by splitting equation~\eqref{eq:delta_potential} into
\begin{align}\label{eq:delta_potential_split}
    \de V(r) & = \frac{1}{\De}\sum_{k= - K}^{4} \al^{(k)} \left(\frac{r}{r_+}\right)^k \notag \\
    & = \frac{1}{\De} \sum_{k=0}^{4} \al^{(k)} \left(\frac{r}{r_+}\right)^k + \frac{1}{\De} \sum_{k= 1}^{K} \al^{(-k)} \left(\frac{r_+}{r}\right)^k
\end{align}
For $k\geq1$, the first generic term in $k$ of the sum can be rewritten as
\begin{align}
    \frac{\al^{(k)}}{\De} \left(\frac{r}{r_+}\right)^k & = \frac{\al^{(k)}}{\De} \left(\frac{r^k - r_+^k}{r_+^k} + 1 \right) \notag \\
    & = \al^{(k)} \left[ \frac{1}{\De} +  \frac{1}{r_+(r-r_-)} \sum_{j = 0}^{k-1} \left( \frac{r}{r_+} \right)^j  \right]
\end{align}
For $k\geq2$ we can further simplify this term as
\begin{widetext}
\begin{align}
    \frac{\al^{(k)}}{\De}  \left(\frac{r}{r_+}\right)^k & = \al^{(k)} \left[ \frac{1}{\De} +  \frac{1}{r_+(r-r_-)} \sum_{j = 0}^{k-1}  \frac{r^j - r_-^j + r_-^j}{r_+^j}   \right] \notag \\
    & = \al^{(k)} \left[ \frac{1}{\De} +  \frac{1}{r_+(r-r_-)} \sum_{j=0}^{k-1} \left( \frac{r_-}{r_+} \right)^j +  \frac{1}{r_- r_+}\sum_{j = 1}^{k-1} \left(\frac{r_-}{r_+} \right)^j \sum_{n=0}^{j-1} \left( \frac{r}{r_-} \right)^n   \right] \notag \\
    & = \al^{(k)} \left[ \frac{1}{\De} +  \frac{1}{r_+(r-r_-)} \sum_{j=0}^{k-1} \left( \frac{r_-}{r_+} \right)^j +  \frac{1}{r_+^2}\sum_{j = 0}^{k-2} \left(\frac{r}{r_-} \right)^j \sum_{n=j}^{k-2} \left( \frac{r_-}{r_+} \right)^n   \right] 
\end{align}
where we obtained the last line by expanding the series and collecting the terms in $r$ to the same power. Summing over all the non-negative values of $k$ yields
\begin{align}
    \frac{1}{\De} \sum_{k=0}^{4} \al^{(k)} \left(\frac{r}{r_+}\right)^k = 
    \frac{1}{\De} \sum_{k=0}^4 \al^{(k)}
    +  \frac{1}{r_+(r-r_-)} \sum_{k=1}^{4} \al^{(k)} \sum_{j=0}^{k-1} \left( \frac{r_-}{r_+} \right)^j 
    + \frac{1}{r_+^2} \sum_{k=0}^2 \left( \frac{r}{r_+} \right)^k \sum_{j=k}^2 \al^{(j+2)} \sum_{n=0}^j \left(\frac{r_-}{r_+} \right)^n
\end{align}
Now, we can perform a mapping between the coefficients $\al^{(k)}$ of equation~\eqref{eq:delta_potential} and the coefficients $A^{(0)}$, $A^{(1)}$ and $\widetilde{\al}^{(k)}$ introduced in equations~\eqref{eq:delta_potential_2}. From a direct comparison we have
\begin{align}
    A^{(0)} & = \sum_{k=0}^4 \al^{(k)} \\
    A^{(1)} & = \sum_{k=1}^{4} \al^{(k)} \sum_{j=0}^{k-1} \left( \frac{r_-}{r_+} \right)^j \\
    \widetilde{\al}^{(k)} & = 
        \sum_{j=k}^2 \al^{(j+2)} \sum_{n=0}^j \left(\frac{r_-}{r_+} \right)^n 
\end{align}

\section{Diagonal quadratic coefficients}\label{app:quadratic}

We show here how to compute the quadratic diagonal coefficients, defined from the next-to-leading-order expansion
\begin{equation}\label{eq:dOmega_dB_quad}
\begin{split}
    \omega & \simeq \omega^0 + \sum_{k} d_{\om}^{(k)} \al^{(k)} + \frac{1}{2} e_{\om}^{(k)} {\al^{(k)}}^2 \,, \\
    B & \simeq B^0 + \sum_{k} d_{B}^{(k)} \al^{(k)} + \frac{1}{2} e_{B}^{(k)} {\al^{(k)}}^2 \,.
\end{split}
\end{equation}
By extending the Taylor expansion~\eqref{eq:Taylor} to the second order in $\al$, we obtain
\begin{equation}\label{eq:Taylor_second}
    \left.\mathcal{L}_j\right|_{\note{GR}} + \al \left.\frac{\dd \mathcal{L}_j}{\dd \al}\right|_\note{GR}  + \frac{\al^2}{2} \left.\frac{\dd^2 \mathcal{L}_j}{\dd \al^2}\right|_\note{GR}  + \Ord(\al)^3 = 0 \,.
\end{equation}
By expanding with the chain rule the total derivative $\dd^2 \mathcal{L}_j/\dd \al^2|_\note{GR}$, one can read off the quadratic coefficients as
\begin{equation}
\begin{split}
    e_\om =& \left(\frac{\partial \mathcal{L}_r}{\partial B}\frac{\partial \mathcal{L}_\th}{\partial \om}-\frac{\partial \mathcal{L}_r}{\partial\om}\frac{\partial \mathcal{L}_\th}{\partial B}\right)^{-1} \Bigg[\frac{\partial^2 \mathcal{L}_r}{\partial\alpha^2}\frac{\partial \mathcal{L}_\th}{\partial B} +2 d_\om\,\frac{\partial^2 \mathcal{L}_r}{\partial\alpha\partial\om}\,\frac{\partial \mathcal{L}_\theta}{\partial B}+2 d_B\,\frac{\partial^2 \mathcal{L}_r}{\partial\alpha\partial B}\,\frac{\partial \mathcal{L}_\theta}{\partial B} \\
    &-d_\om^2\,\left(\frac{\partial^2 \mathcal{L}_\th}{\partial\om^2}\frac{\partial \mathcal{L}_r}{\partial B}-\frac{\partial^2\mathcal{L}_r}{\partial\om^2}\frac{\partial \mathcal{L}_\th}{\partial B}\right)-d_B^2\,\left(\frac{\partial^2 \mathcal{L}_\th}{\partial B^2}\frac{\partial \mathcal{L}_r}{\partial B}-\frac{\partial^2\mathcal{L}_r}{\partial B^2}\frac{\partial \mathcal{L}_\th}{\partial B}\right)+2d_\om\,d_B\,\left(\frac{\partial^2\mathcal{L}_r}{\partial \om \partial B}\frac{\partial \mathcal{L}_\th}{\partial B}-\frac{\partial^2\mathcal{L}_\th}{\partial \om \partial B}\frac{\partial \mathcal{L}_r}{\partial B}\right)\Bigg]\,, \\    
    e_B =&- \left(\frac{\partial \mathcal{L}_r}{\partial B}\frac{\partial \mathcal{L}_\th}{\partial \om}-\frac{\partial \mathcal{L}_r}{\partial\om}\frac{\partial \mathcal{L}_\th}{\partial B}\right)^{-1}\Bigg[\frac{\partial^2 \mathcal{L}_r}{\partial\alpha^2}\frac{\partial \mathcal{L}_\th}{\partial \om}+2 d_\om\,\frac{\partial^2 \mathcal{L}_r}{\partial\alpha\partial\om}\,\frac{\partial \mathcal{L}_\theta}{\partial \om}+2 d_B\,\frac{\partial^2 \mathcal{L}_r}{\partial\alpha\partial B}\,\frac{\partial \mathcal{L}_\theta}{\partial \om} \\
    &-d_\om^2\,\left(\frac{\partial^2 \mathcal{L}_r}{\partial\om^2}\frac{\partial \mathcal{L}_\th}{\partial \om}-\frac{\partial^2\mathcal{L}_\th}{\partial\om^2}\frac{\partial \mathcal{L}_r}{\partial \om}\right)-d_B^2\,\left(\frac{\partial^2 \mathcal{L}_r}{\partial B^2}\frac{\partial \mathcal{L}_\th}{\partial \om}-\frac{\partial^2\mathcal{L}_\th}{\partial B^2}\frac{\partial \mathcal{L}_r}{\partial \om}\right) +2d_\om\,d_B\,\left(\frac{\partial^2\mathcal{L}_r}{\partial \om \partial B}\frac{\partial \mathcal{L}_\th}{\partial \om}-\frac{\partial^2\mathcal{L}_\th}{\partial \om \partial B}\frac{\partial \mathcal{L}_r}{\partial \om}\right)\Bigg]\,,
\end{split}
\end{equation}

\end{widetext}

\bibliography{literature}

\begin{thebibliography}{69}%
\makeatletter
\providecommand \@ifxundefined [1]{%
 \@ifx{#1\undefined}
}%
\providecommand \@ifnum [1]{%
 \ifnum #1\expandafter \@firstoftwo
 \else \expandafter \@secondoftwo
 \fi
}%
\providecommand \@ifx [1]{%
 \ifx #1\expandafter \@firstoftwo
 \else \expandafter \@secondoftwo
 \fi
}%
\providecommand \natexlab [1]{#1}%
\providecommand \enquote  [1]{``#1''}%
\providecommand \bibnamefont  [1]{#1}%
\providecommand \bibfnamefont [1]{#1}%
\providecommand \citenamefont [1]{#1}%
\providecommand \href@noop [0]{\@secondoftwo}%
\providecommand \href [0]{\begingroup \@sanitize@url \@href}%
\providecommand \@href[1]{\@@startlink{#1}\@@href}%
\providecommand \@@href[1]{\endgroup#1\@@endlink}%
\providecommand \@sanitize@url [0]{\catcode `\\12\catcode `\$12\catcode
  `\&12\catcode `\#12\catcode `\^12\catcode `\_12\catcode `\%12\relax}%
\providecommand \@@startlink[1]{}%
\providecommand \@@endlink[0]{}%
\providecommand \url  [0]{\begingroup\@sanitize@url \@url }%
\providecommand \@url [1]{\endgroup\@href {#1}{\urlprefix }}%
\providecommand \urlprefix  [0]{URL }%
\providecommand \Eprint [0]{\href }%
\providecommand \doibase [0]{https://doi.org/}%
\providecommand \selectlanguage [0]{\@gobble}%
\providecommand \bibinfo  [0]{\@secondoftwo}%
\providecommand \bibfield  [0]{\@secondoftwo}%
\providecommand \translation [1]{[#1]}%
\providecommand \BibitemOpen [0]{}%
\providecommand \bibitemStop [0]{}%
\providecommand \bibitemNoStop [0]{.\EOS\space}%
\providecommand \EOS [0]{\spacefactor3000\relax}%
\providecommand \BibitemShut  [1]{\csname bibitem#1\endcsname}%
\let\auto@bib@innerbib\@empty
\bibitem [{\citenamefont {Abbott}\ \emph {et~al.}(2023)\citenamefont {Abbott}
  \emph {et~al.}}]{KAGRA:2021vkt}%
  \BibitemOpen
  \bibfield  {author} {\bibinfo {author} {\bibfnamefont {R.}~\bibnamefont
  {Abbott}} \emph {et~al.} (\bibinfo {collaboration} {KAGRA, VIRGO, LIGO
  Scientific}),\ }\bibfield  {title} {\bibinfo {title} {{GWTC-3: Compact Binary
  Coalescences Observed by LIGO and Virgo during the Second Part of the Third
  Observing Run}},\ }\href {https://doi.org/10.1103/PhysRevX.13.041039}
  {\bibfield  {journal} {\bibinfo  {journal} {Phys. Rev. X}\ }\textbf {\bibinfo
  {volume} {13}},\ \bibinfo {pages} {041039} (\bibinfo {year} {2023})},\
  \Eprint {https://arxiv.org/abs/2111.03606} {arXiv:2111.03606 [gr-qc]}
  \BibitemShut {NoStop}%
\bibitem [{\citenamefont {Kokkotas}\ and\ \citenamefont
  {Schmidt}(1999)}]{Kokkotas:1999bd}%
  \BibitemOpen
  \bibfield  {author} {\bibinfo {author} {\bibfnamefont {K.~D.}\ \bibnamefont
  {Kokkotas}}\ and\ \bibinfo {author} {\bibfnamefont {B.~G.}\ \bibnamefont
  {Schmidt}},\ }\bibfield  {title} {\bibinfo {title} {{Quasinormal modes of
  stars and black holes}},\ }\href {https://doi.org/10.12942/lrr-1999-2}
  {\bibfield  {journal} {\bibinfo  {journal} {Living Rev. Rel.}\ }\textbf
  {\bibinfo {volume} {2}},\ \bibinfo {pages} {2} (\bibinfo {year} {1999})},\
  \Eprint {https://arxiv.org/abs/gr-qc/9909058} {arXiv:gr-qc/9909058}
  \BibitemShut {NoStop}%
\bibitem [{\citenamefont {Berti}\ \emph {et~al.}(2009)\citenamefont {Berti},
  \citenamefont {Cardoso},\ and\ \citenamefont {Starinets}}]{Berti:2009kk}%
  \BibitemOpen
  \bibfield  {author} {\bibinfo {author} {\bibfnamefont {E.}~\bibnamefont
  {Berti}}, \bibinfo {author} {\bibfnamefont {V.}~\bibnamefont {Cardoso}},\
  and\ \bibinfo {author} {\bibfnamefont {A.~O.}\ \bibnamefont {Starinets}},\
  }\bibfield  {title} {\bibinfo {title} {{Quasinormal modes of black holes and
  black branes}},\ }\href {https://doi.org/10.1088/0264-9381/26/16/163001}
  {\bibfield  {journal} {\bibinfo  {journal} {Class. Quant. Grav.}\ }\textbf
  {\bibinfo {volume} {26}},\ \bibinfo {pages} {163001} (\bibinfo {year}
  {2009})},\ \Eprint {https://arxiv.org/abs/0905.2975} {arXiv:0905.2975
  [gr-qc]} \BibitemShut {NoStop}%
\bibitem [{\citenamefont {Franchini}\ and\ \citenamefont
  {V\"olkel}(2023{\natexlab{a}})}]{Franchini:2023eda}%
  \BibitemOpen
  \bibfield  {author} {\bibinfo {author} {\bibfnamefont {N.}~\bibnamefont
  {Franchini}}\ and\ \bibinfo {author} {\bibfnamefont {S.~H.}\ \bibnamefont
  {V\"olkel}},\ }\bibfield  {title} {\bibinfo {title} {{Testing General
  Relativity with Black Hole Quasi-Normal Modes}}\ }\href@noop {} {} (\bibinfo
  {year} {2023}{\natexlab{a}}),\ \Eprint {https://arxiv.org/abs/2305.01696}
  {arXiv:2305.01696 [gr-qc]} \BibitemShut {NoStop}%
\bibitem [{\citenamefont {Abbott}\ \emph {et~al.}(2016)\citenamefont {Abbott}
  \emph {et~al.}}]{LIGOScientific:2016lio}%
  \BibitemOpen
  \bibfield  {author} {\bibinfo {author} {\bibfnamefont {B.~P.}\ \bibnamefont
  {Abbott}} \emph {et~al.} (\bibinfo {collaboration} {LIGO Scientific,
  Virgo}),\ }\bibfield  {title} {\bibinfo {title} {{Tests of general relativity
  with GW150914}},\ }\href {https://doi.org/10.1103/PhysRevLett.116.221101}
  {\bibfield  {journal} {\bibinfo  {journal} {Phys. Rev. Lett.}\ }\textbf
  {\bibinfo {volume} {116}},\ \bibinfo {pages} {221101} (\bibinfo {year}
  {2016})},\ \bibinfo {note} {[Erratum: Phys.Rev.Lett. 121, 129902 (2018)]},\
  \Eprint {https://arxiv.org/abs/1602.03841} {arXiv:1602.03841 [gr-qc]}
  \BibitemShut {NoStop}%
\bibitem [{\citenamefont {Abbott}\ \emph
  {et~al.}(2021{\natexlab{a}})\citenamefont {Abbott} \emph
  {et~al.}}]{LIGOScientific:2020tif}%
  \BibitemOpen
  \bibfield  {author} {\bibinfo {author} {\bibfnamefont {R.}~\bibnamefont
  {Abbott}} \emph {et~al.} (\bibinfo {collaboration} {LIGO Scientific,
  Virgo}),\ }\bibfield  {title} {\bibinfo {title} {{Tests of general relativity
  with binary black holes from the second LIGO-Virgo gravitational-wave
  transient catalog}},\ }\href {https://doi.org/10.1103/PhysRevD.103.122002}
  {\bibfield  {journal} {\bibinfo  {journal} {Phys. Rev. D}\ }\textbf {\bibinfo
  {volume} {103}},\ \bibinfo {pages} {122002} (\bibinfo {year}
  {2021}{\natexlab{a}})},\ \Eprint {https://arxiv.org/abs/2010.14529}
  {arXiv:2010.14529 [gr-qc]} \BibitemShut {NoStop}%
\bibitem [{\citenamefont {Abbott}\ \emph
  {et~al.}(2021{\natexlab{b}})\citenamefont {Abbott} \emph
  {et~al.}}]{LIGOScientific:2021sio}%
  \BibitemOpen
  \bibfield  {author} {\bibinfo {author} {\bibfnamefont {R.}~\bibnamefont
  {Abbott}} \emph {et~al.} (\bibinfo {collaboration} {LIGO Scientific, VIRGO,
  KAGRA}),\ }\bibfield  {title} {\bibinfo {title} {{Tests of General Relativity
  with GWTC-3}}\ }\href@noop {} {} (\bibinfo {year} {2021}{\natexlab{b}}),\
  \Eprint {https://arxiv.org/abs/2112.06861} {arXiv:2112.06861 [gr-qc]}
  \BibitemShut {NoStop}%
\bibitem [{\citenamefont {Carullo}\ \emph {et~al.}(2019)\citenamefont
  {Carullo}, \citenamefont {Del~Pozzo},\ and\ \citenamefont
  {Veitch}}]{Carullo:2019flw}%
  \BibitemOpen
  \bibfield  {author} {\bibinfo {author} {\bibfnamefont {G.}~\bibnamefont
  {Carullo}}, \bibinfo {author} {\bibfnamefont {W.}~\bibnamefont {Del~Pozzo}},\
  and\ \bibinfo {author} {\bibfnamefont {J.}~\bibnamefont {Veitch}},\
  }\bibfield  {title} {\bibinfo {title} {{Observational Black Hole
  Spectroscopy: A time-domain multimode analysis of GW150914}},\ }\href
  {https://doi.org/10.1103/PhysRevD.99.123029} {\bibfield  {journal} {\bibinfo
  {journal} {Phys. Rev. D}\ }\textbf {\bibinfo {volume} {99}},\ \bibinfo
  {pages} {123029} (\bibinfo {year} {2019})},\ \bibinfo {note} {[Erratum:
  Phys.Rev.D 100, 089903 (2019)]},\ \Eprint {https://arxiv.org/abs/1902.07527}
  {arXiv:1902.07527 [gr-qc]} \BibitemShut {NoStop}%
\bibitem [{\citenamefont {Isi}\ \emph {et~al.}(2019)\citenamefont {Isi},
  \citenamefont {Giesler}, \citenamefont {Farr}, \citenamefont {Scheel},\ and\
  \citenamefont {Teukolsky}}]{Isi:2019aib}%
  \BibitemOpen
  \bibfield  {author} {\bibinfo {author} {\bibfnamefont {M.}~\bibnamefont
  {Isi}}, \bibinfo {author} {\bibfnamefont {M.}~\bibnamefont {Giesler}},
  \bibinfo {author} {\bibfnamefont {W.~M.}\ \bibnamefont {Farr}}, \bibinfo
  {author} {\bibfnamefont {M.~A.}\ \bibnamefont {Scheel}},\ and\ \bibinfo
  {author} {\bibfnamefont {S.~A.}\ \bibnamefont {Teukolsky}},\ }\bibfield
  {title} {\bibinfo {title} {{Testing the no-hair theorem with GW150914}},\
  }\href {https://doi.org/10.1103/PhysRevLett.123.111102} {\bibfield  {journal}
  {\bibinfo  {journal} {Phys. Rev. Lett.}\ }\textbf {\bibinfo {volume} {123}},\
  \bibinfo {pages} {111102} (\bibinfo {year} {2019})},\ \Eprint
  {https://arxiv.org/abs/1905.00869} {arXiv:1905.00869 [gr-qc]} \BibitemShut
  {NoStop}%
\bibitem [{\citenamefont {Cotesta}\ \emph {et~al.}(2022)\citenamefont
  {Cotesta}, \citenamefont {Carullo}, \citenamefont {Berti},\ and\
  \citenamefont {Cardoso}}]{Cotesta:2022pci}%
  \BibitemOpen
  \bibfield  {author} {\bibinfo {author} {\bibfnamefont {R.}~\bibnamefont
  {Cotesta}}, \bibinfo {author} {\bibfnamefont {G.}~\bibnamefont {Carullo}},
  \bibinfo {author} {\bibfnamefont {E.}~\bibnamefont {Berti}},\ and\ \bibinfo
  {author} {\bibfnamefont {V.}~\bibnamefont {Cardoso}},\ }\bibfield  {title}
  {\bibinfo {title} {{Analysis of Ringdown Overtones in GW150914}},\ }\href
  {https://doi.org/10.1103/PhysRevLett.129.111102} {\bibfield  {journal}
  {\bibinfo  {journal} {Phys. Rev. Lett.}\ }\textbf {\bibinfo {volume} {129}},\
  \bibinfo {pages} {111102} (\bibinfo {year} {2022})},\ \Eprint
  {https://arxiv.org/abs/2201.00822} {arXiv:2201.00822 [gr-qc]} \BibitemShut
  {NoStop}%
\bibitem [{\citenamefont {Finch}\ and\ \citenamefont
  {Moore}(2022)}]{Finch:2022ynt}%
  \BibitemOpen
  \bibfield  {author} {\bibinfo {author} {\bibfnamefont {E.}~\bibnamefont
  {Finch}}\ and\ \bibinfo {author} {\bibfnamefont {C.~J.}\ \bibnamefont
  {Moore}},\ }\bibfield  {title} {\bibinfo {title} {{Searching for a ringdown
  overtone in GW150914}},\ }\href {https://doi.org/10.1103/PhysRevD.106.043005}
  {\bibfield  {journal} {\bibinfo  {journal} {Phys. Rev. D}\ }\textbf {\bibinfo
  {volume} {106}},\ \bibinfo {pages} {043005} (\bibinfo {year} {2022})},\
  \Eprint {https://arxiv.org/abs/2205.07809} {arXiv:2205.07809 [gr-qc]}
  \BibitemShut {NoStop}%
\bibitem [{\citenamefont {Isi}\ and\ \citenamefont {Farr}(2022)}]{Isi:2022mhy}%
  \BibitemOpen
  \bibfield  {author} {\bibinfo {author} {\bibfnamefont {M.}~\bibnamefont
  {Isi}}\ and\ \bibinfo {author} {\bibfnamefont {W.~M.}\ \bibnamefont {Farr}},\
  }\bibfield  {title} {\bibinfo {title} {{Revisiting the ringdown of
  GW150914}}\ }\href@noop {} {} (\bibinfo {year} {2022}),\ \Eprint
  {https://arxiv.org/abs/2202.02941} {arXiv:2202.02941 [gr-qc]} \BibitemShut
  {NoStop}%
\bibitem [{\citenamefont {Carullo}\ \emph {et~al.}(2023)\citenamefont
  {Carullo}, \citenamefont {Cotesta}, \citenamefont {Berti},\ and\
  \citenamefont {Cardoso}}]{Carullo:2023gtf}%
  \BibitemOpen
  \bibfield  {author} {\bibinfo {author} {\bibfnamefont {G.}~\bibnamefont
  {Carullo}}, \bibinfo {author} {\bibfnamefont {R.}~\bibnamefont {Cotesta}},
  \bibinfo {author} {\bibfnamefont {E.}~\bibnamefont {Berti}},\ and\ \bibinfo
  {author} {\bibfnamefont {V.}~\bibnamefont {Cardoso}},\ }\bibfield  {title}
  {\bibinfo {title} {{Reply to Comment on ''Analysis of Ringdown Overtones in
  GW150914''}},\ }\href {https://doi.org/10.1103/PhysRevLett.131.169002}
  {\bibfield  {journal} {\bibinfo  {journal} {Phys. Rev. Lett.}\ }\textbf
  {\bibinfo {volume} {131}},\ \bibinfo {pages} {169002} (\bibinfo {year}
  {2023})},\ \Eprint {https://arxiv.org/abs/2310.20625} {arXiv:2310.20625
  [gr-qc]} \BibitemShut {NoStop}%
\bibitem [{\citenamefont {Crisostomi}\ \emph {et~al.}(2023)\citenamefont
  {Crisostomi}, \citenamefont {Dey}, \citenamefont {Barausse},\ and\
  \citenamefont {Trotta}}]{Crisostomi:2023tle}%
  \BibitemOpen
  \bibfield  {author} {\bibinfo {author} {\bibfnamefont {M.}~\bibnamefont
  {Crisostomi}}, \bibinfo {author} {\bibfnamefont {K.}~\bibnamefont {Dey}},
  \bibinfo {author} {\bibfnamefont {E.}~\bibnamefont {Barausse}},\ and\
  \bibinfo {author} {\bibfnamefont {R.}~\bibnamefont {Trotta}},\ }\bibfield
  {title} {\bibinfo {title} {{Neural posterior estimation with guaranteed exact
  coverage: The ringdown of GW150914}},\ }\href
  {https://doi.org/10.1103/PhysRevD.108.044029} {\bibfield  {journal} {\bibinfo
   {journal} {Phys. Rev. D}\ }\textbf {\bibinfo {volume} {108}},\ \bibinfo
  {pages} {044029} (\bibinfo {year} {2023})},\ \Eprint
  {https://arxiv.org/abs/2305.18528} {arXiv:2305.18528 [gr-qc]} \BibitemShut
  {NoStop}%
\bibitem [{\citenamefont {Pacilio}\ \emph {et~al.}(2024)\citenamefont
  {Pacilio}, \citenamefont {Bhagwat},\ and\ \citenamefont
  {Cotesta}}]{Pacilio:2024qcq}%
  \BibitemOpen
  \bibfield  {author} {\bibinfo {author} {\bibfnamefont {C.}~\bibnamefont
  {Pacilio}}, \bibinfo {author} {\bibfnamefont {S.}~\bibnamefont {Bhagwat}},\
  and\ \bibinfo {author} {\bibfnamefont {R.}~\bibnamefont {Cotesta}},\
  }\bibfield  {title} {\bibinfo {title} {{Simulation-based inference of black
  hole ringdowns in the time domain}}\ }\href@noop {} {} (\bibinfo {year}
  {2024}),\ \Eprint {https://arxiv.org/abs/2404.11373} {arXiv:2404.11373
  [gr-qc]} \BibitemShut {NoStop}%
\bibitem [{\citenamefont {Gennari}\ \emph {et~al.}(2024)\citenamefont
  {Gennari}, \citenamefont {Carullo},\ and\ \citenamefont
  {Del~Pozzo}}]{Gennari:2023gmx}%
  \BibitemOpen
  \bibfield  {author} {\bibinfo {author} {\bibfnamefont {V.}~\bibnamefont
  {Gennari}}, \bibinfo {author} {\bibfnamefont {G.}~\bibnamefont {Carullo}},\
  and\ \bibinfo {author} {\bibfnamefont {W.}~\bibnamefont {Del~Pozzo}},\
  }\bibfield  {title} {\bibinfo {title} {{Searching for ringdown higher modes
  with a numerical relativity-informed post-merger model}},\ }\href
  {https://doi.org/10.1140/epjc/s10052-024-12550-x} {\bibfield  {journal}
  {\bibinfo  {journal} {Eur. Phys. J. C}\ }\textbf {\bibinfo {volume} {84}},\
  \bibinfo {pages} {233} (\bibinfo {year} {2024})},\ \Eprint
  {https://arxiv.org/abs/2312.12515} {arXiv:2312.12515 [gr-qc]} \BibitemShut
  {NoStop}%
\bibitem [{\citenamefont {Maselli}\ \emph {et~al.}(2020)\citenamefont
  {Maselli}, \citenamefont {Pani}, \citenamefont {Gualtieri},\ and\
  \citenamefont {Berti}}]{Maselli:2019mjd}%
  \BibitemOpen
  \bibfield  {author} {\bibinfo {author} {\bibfnamefont {A.}~\bibnamefont
  {Maselli}}, \bibinfo {author} {\bibfnamefont {P.}~\bibnamefont {Pani}},
  \bibinfo {author} {\bibfnamefont {L.}~\bibnamefont {Gualtieri}},\ and\
  \bibinfo {author} {\bibfnamefont {E.}~\bibnamefont {Berti}},\ }\bibfield
  {title} {\bibinfo {title} {{Parametrized ringdown spin expansion
  coefficients: a data-analysis framework for black-hole spectroscopy with
  multiple events}},\ }\href {https://doi.org/10.1103/PhysRevD.101.024043}
  {\bibfield  {journal} {\bibinfo  {journal} {Phys. Rev. D}\ }\textbf {\bibinfo
  {volume} {101}},\ \bibinfo {pages} {024043} (\bibinfo {year} {2020})},\
  \Eprint {https://arxiv.org/abs/1910.12893} {arXiv:1910.12893 [gr-qc]}
  \BibitemShut {NoStop}%
\bibitem [{\citenamefont {Carullo}(2021)}]{Carullo:2021dui}%
  \BibitemOpen
  \bibfield  {author} {\bibinfo {author} {\bibfnamefont {G.}~\bibnamefont
  {Carullo}},\ }\bibfield  {title} {\bibinfo {title} {{Enhancing modified
  gravity detection from gravitational-wave observations using the parametrized
  ringdown spin expansion coeffcients formalism}},\ }\href
  {https://doi.org/10.1103/PhysRevD.103.124043} {\bibfield  {journal} {\bibinfo
   {journal} {Phys. Rev. D}\ }\textbf {\bibinfo {volume} {103}},\ \bibinfo
  {pages} {124043} (\bibinfo {year} {2021})},\ \Eprint
  {https://arxiv.org/abs/2102.05939} {arXiv:2102.05939 [gr-qc]} \BibitemShut
  {NoStop}%
\bibitem [{\citenamefont {Maselli}\ \emph {et~al.}(2024)\citenamefont
  {Maselli}, \citenamefont {Yi}, \citenamefont {Pierini}, \citenamefont
  {Vellucci}, \citenamefont {Reali}, \citenamefont {Gualtieri},\ and\
  \citenamefont {Berti}}]{Maselli:2023khq}%
  \BibitemOpen
  \bibfield  {author} {\bibinfo {author} {\bibfnamefont {A.}~\bibnamefont
  {Maselli}}, \bibinfo {author} {\bibfnamefont {S.}~\bibnamefont {Yi}},
  \bibinfo {author} {\bibfnamefont {L.}~\bibnamefont {Pierini}}, \bibinfo
  {author} {\bibfnamefont {V.}~\bibnamefont {Vellucci}}, \bibinfo {author}
  {\bibfnamefont {L.}~\bibnamefont {Reali}}, \bibinfo {author} {\bibfnamefont
  {L.}~\bibnamefont {Gualtieri}},\ and\ \bibinfo {author} {\bibfnamefont
  {E.}~\bibnamefont {Berti}},\ }\bibfield  {title} {\bibinfo {title} {{Black
  hole spectroscopy beyond Kerr: Agnostic and theory-based tests with
  next-generation interferometers}},\ }\href
  {https://doi.org/10.1103/PhysRevD.109.064060} {\bibfield  {journal} {\bibinfo
   {journal} {Phys. Rev. D}\ }\textbf {\bibinfo {volume} {109}},\ \bibinfo
  {pages} {064060} (\bibinfo {year} {2024})},\ \Eprint
  {https://arxiv.org/abs/2311.14803} {arXiv:2311.14803 [gr-qc]} \BibitemShut
  {NoStop}%
\bibitem [{\citenamefont {Carullo}\ \emph {et~al.}(2022)\citenamefont
  {Carullo}, \citenamefont {Laghi}, \citenamefont {Johnson-McDaniel},
  \citenamefont {Del~Pozzo}, \citenamefont {Dias}, \citenamefont {Godazgar},\
  and\ \citenamefont {Santos}}]{Carullo:2021oxn}%
  \BibitemOpen
  \bibfield  {author} {\bibinfo {author} {\bibfnamefont {G.}~\bibnamefont
  {Carullo}}, \bibinfo {author} {\bibfnamefont {D.}~\bibnamefont {Laghi}},
  \bibinfo {author} {\bibfnamefont {N.~K.}\ \bibnamefont {Johnson-McDaniel}},
  \bibinfo {author} {\bibfnamefont {W.}~\bibnamefont {Del~Pozzo}}, \bibinfo
  {author} {\bibfnamefont {O.~J.~C.}\ \bibnamefont {Dias}}, \bibinfo {author}
  {\bibfnamefont {M.}~\bibnamefont {Godazgar}},\ and\ \bibinfo {author}
  {\bibfnamefont {J.~E.}\ \bibnamefont {Santos}},\ }\bibfield  {title}
  {\bibinfo {title} {{Constraints on Kerr-Newman black holes from
  merger-ringdown gravitational-wave observations}},\ }\href
  {https://doi.org/10.1103/PhysRevD.105.062009} {\bibfield  {journal} {\bibinfo
   {journal} {Phys. Rev. D}\ }\textbf {\bibinfo {volume} {105}},\ \bibinfo
  {pages} {062009} (\bibinfo {year} {2022})},\ \Eprint
  {https://arxiv.org/abs/2109.13961} {arXiv:2109.13961 [gr-qc]} \BibitemShut
  {NoStop}%
\bibitem [{\citenamefont {Pani}\ and\ \citenamefont
  {Cardoso}(2009)}]{Pani:2009wy}%
  \BibitemOpen
  \bibfield  {author} {\bibinfo {author} {\bibfnamefont {P.}~\bibnamefont
  {Pani}}\ and\ \bibinfo {author} {\bibfnamefont {V.}~\bibnamefont {Cardoso}},\
  }\bibfield  {title} {\bibinfo {title} {{Are black holes in alternative
  theories serious astrophysical candidates? The Case for
  Einstein-Dilaton-Gauss-Bonnet black holes}},\ }\href
  {https://doi.org/10.1103/PhysRevD.79.084031} {\bibfield  {journal} {\bibinfo
  {journal} {Phys. Rev. D}\ }\textbf {\bibinfo {volume} {79}},\ \bibinfo
  {pages} {084031} (\bibinfo {year} {2009})},\ \Eprint
  {https://arxiv.org/abs/0902.1569} {arXiv:0902.1569 [gr-qc]} \BibitemShut
  {NoStop}%
\bibitem [{\citenamefont {Yunes}\ and\ \citenamefont
  {Pretorius}(2009)}]{Yunes:2009hc}%
  \BibitemOpen
  \bibfield  {author} {\bibinfo {author} {\bibfnamefont {N.}~\bibnamefont
  {Yunes}}\ and\ \bibinfo {author} {\bibfnamefont {F.}~\bibnamefont
  {Pretorius}},\ }\bibfield  {title} {\bibinfo {title} {{Dynamical Chern-Simons
  Modified Gravity. I. Spinning Black Holes in the Slow-Rotation
  Approximation}},\ }\href {https://doi.org/10.1103/PhysRevD.79.084043}
  {\bibfield  {journal} {\bibinfo  {journal} {Phys. Rev. D}\ }\textbf {\bibinfo
  {volume} {79}},\ \bibinfo {pages} {084043} (\bibinfo {year} {2009})},\
  \Eprint {https://arxiv.org/abs/0902.4669} {arXiv:0902.4669 [gr-qc]}
  \BibitemShut {NoStop}%
\bibitem [{\citenamefont {Pani}\ \emph {et~al.}(2011)\citenamefont {Pani},
  \citenamefont {Macedo}, \citenamefont {Crispino},\ and\ \citenamefont
  {Cardoso}}]{Pani:2011gy}%
  \BibitemOpen
  \bibfield  {author} {\bibinfo {author} {\bibfnamefont {P.}~\bibnamefont
  {Pani}}, \bibinfo {author} {\bibfnamefont {C.~F.~B.}\ \bibnamefont {Macedo}},
  \bibinfo {author} {\bibfnamefont {L.~C.~B.}\ \bibnamefont {Crispino}},\ and\
  \bibinfo {author} {\bibfnamefont {V.}~\bibnamefont {Cardoso}},\ }\bibfield
  {title} {\bibinfo {title} {{Slowly rotating black holes in alternative
  theories of gravity}},\ }\href {https://doi.org/10.1103/PhysRevD.84.087501}
  {\bibfield  {journal} {\bibinfo  {journal} {Phys. Rev. D}\ }\textbf {\bibinfo
  {volume} {84}},\ \bibinfo {pages} {087501} (\bibinfo {year} {2011})},\
  \Eprint {https://arxiv.org/abs/1109.3996} {arXiv:1109.3996 [gr-qc]}
  \BibitemShut {NoStop}%
\bibitem [{\citenamefont {Maselli}\ \emph {et~al.}(2015)\citenamefont
  {Maselli}, \citenamefont {Pani}, \citenamefont {Gualtieri},\ and\
  \citenamefont {Ferrari}}]{Maselli:2015tta}%
  \BibitemOpen
  \bibfield  {author} {\bibinfo {author} {\bibfnamefont {A.}~\bibnamefont
  {Maselli}}, \bibinfo {author} {\bibfnamefont {P.}~\bibnamefont {Pani}},
  \bibinfo {author} {\bibfnamefont {L.}~\bibnamefont {Gualtieri}},\ and\
  \bibinfo {author} {\bibfnamefont {V.}~\bibnamefont {Ferrari}},\ }\bibfield
  {title} {\bibinfo {title} {{Rotating black holes in
  Einstein-Dilaton-Gauss-Bonnet gravity with finite coupling}},\ }\href
  {https://doi.org/10.1103/PhysRevD.92.083014} {\bibfield  {journal} {\bibinfo
  {journal} {Phys. Rev. D}\ }\textbf {\bibinfo {volume} {92}},\ \bibinfo
  {pages} {083014} (\bibinfo {year} {2015})},\ \Eprint
  {https://arxiv.org/abs/1507.00680} {arXiv:1507.00680 [gr-qc]} \BibitemShut
  {NoStop}%
\bibitem [{\citenamefont {Cano}\ and\ \citenamefont
  {Ruip\'erez}(2019)}]{Cano:2019ore}%
  \BibitemOpen
  \bibfield  {author} {\bibinfo {author} {\bibfnamefont {P.~A.}\ \bibnamefont
  {Cano}}\ and\ \bibinfo {author} {\bibfnamefont {A.}~\bibnamefont
  {Ruip\'erez}},\ }\bibfield  {title} {\bibinfo {title} {{Leading
  higher-derivative corrections to Kerr geometry}},\ }\href
  {https://doi.org/10.1007/JHEP05(2019)189} {\bibfield  {journal} {\bibinfo
  {journal} {JHEP}\ }\textbf {\bibinfo {volume} {05}},\ \bibinfo {pages}
  {189}},\ \bibinfo {note} {[Erratum: JHEP 03, 187 (2020)]},\ \Eprint
  {https://arxiv.org/abs/1901.01315} {arXiv:1901.01315 [gr-qc]} \BibitemShut
  {NoStop}%
\bibitem [{\citenamefont {Pani}(2013)}]{Pani:2013pma}%
  \BibitemOpen
  \bibfield  {author} {\bibinfo {author} {\bibfnamefont {P.}~\bibnamefont
  {Pani}},\ }\bibfield  {title} {\bibinfo {title} {{Advanced Methods in
  Black-Hole Perturbation Theory}},\ }\href
  {https://doi.org/10.1142/S0217751X13400186} {\bibfield  {journal} {\bibinfo
  {journal} {Int. J. Mod. Phys. A}\ }\textbf {\bibinfo {volume} {28}},\
  \bibinfo {pages} {1340018} (\bibinfo {year} {2013})},\ \Eprint
  {https://arxiv.org/abs/1305.6759} {arXiv:1305.6759 [gr-qc]} \BibitemShut
  {NoStop}%
\bibitem [{\citenamefont {Franchini}(2023)}]{Franchini:2023xhd}%
  \BibitemOpen
  \bibfield  {author} {\bibinfo {author} {\bibfnamefont {N.}~\bibnamefont
  {Franchini}},\ }\bibfield  {title} {\bibinfo {title} {{Slow rotation black
  hole perturbation theory}},\ }\href
  {https://doi.org/10.1103/PhysRevD.108.044079} {\bibfield  {journal} {\bibinfo
   {journal} {Phys. Rev. D}\ }\textbf {\bibinfo {volume} {108}},\ \bibinfo
  {pages} {044079} (\bibinfo {year} {2023})},\ \Eprint
  {https://arxiv.org/abs/2305.19313} {arXiv:2305.19313 [gr-qc]} \BibitemShut
  {NoStop}%
\bibitem [{\citenamefont {Cano}\ \emph {et~al.}(2022)\citenamefont {Cano},
  \citenamefont {Fransen}, \citenamefont {Hertog},\ and\ \citenamefont
  {Maenaut}}]{Cano:2021myl}%
  \BibitemOpen
  \bibfield  {author} {\bibinfo {author} {\bibfnamefont {P.~A.}\ \bibnamefont
  {Cano}}, \bibinfo {author} {\bibfnamefont {K.}~\bibnamefont {Fransen}},
  \bibinfo {author} {\bibfnamefont {T.}~\bibnamefont {Hertog}},\ and\ \bibinfo
  {author} {\bibfnamefont {S.}~\bibnamefont {Maenaut}},\ }\bibfield  {title}
  {\bibinfo {title} {{Gravitational ringing of rotating black holes in
  higher-derivative gravity}},\ }\href
  {https://doi.org/10.1103/PhysRevD.105.024064} {\bibfield  {journal} {\bibinfo
   {journal} {Phys. Rev. D}\ }\textbf {\bibinfo {volume} {105}},\ \bibinfo
  {pages} {024064} (\bibinfo {year} {2022})},\ \Eprint
  {https://arxiv.org/abs/2110.11378} {arXiv:2110.11378 [gr-qc]} \BibitemShut
  {NoStop}%
\bibitem [{\citenamefont {Pierini}\ and\ \citenamefont
  {Gualtieri}(2021)}]{Pierini:2021jxd}%
  \BibitemOpen
  \bibfield  {author} {\bibinfo {author} {\bibfnamefont {L.}~\bibnamefont
  {Pierini}}\ and\ \bibinfo {author} {\bibfnamefont {L.}~\bibnamefont
  {Gualtieri}},\ }\bibfield  {title} {\bibinfo {title} {{Quasi-normal modes of
  rotating black holes in Einstein-dilaton Gauss-Bonnet gravity: the first
  order in rotation}},\ }\href {https://doi.org/10.1103/PhysRevD.103.124017}
  {\bibfield  {journal} {\bibinfo  {journal} {Phys. Rev. D}\ }\textbf {\bibinfo
  {volume} {103}},\ \bibinfo {pages} {124017} (\bibinfo {year} {2021})},\
  \Eprint {https://arxiv.org/abs/2103.09870} {arXiv:2103.09870 [gr-qc]}
  \BibitemShut {NoStop}%
\bibitem [{\citenamefont {Pierini}\ and\ \citenamefont
  {Gualtieri}(2022)}]{Pierini:2022eim}%
  \BibitemOpen
  \bibfield  {author} {\bibinfo {author} {\bibfnamefont {L.}~\bibnamefont
  {Pierini}}\ and\ \bibinfo {author} {\bibfnamefont {L.}~\bibnamefont
  {Gualtieri}},\ }\bibfield  {title} {\bibinfo {title} {{Quasinormal modes of
  rotating black holes in Einstein-dilaton Gauss-Bonnet gravity: The second
  order in rotation}},\ }\href {https://doi.org/10.1103/PhysRevD.106.104009}
  {\bibfield  {journal} {\bibinfo  {journal} {Phys. Rev. D}\ }\textbf {\bibinfo
  {volume} {106}},\ \bibinfo {pages} {104009} (\bibinfo {year} {2022})},\
  \Eprint {https://arxiv.org/abs/2207.11267} {arXiv:2207.11267 [gr-qc]}
  \BibitemShut {NoStop}%
\bibitem [{\citenamefont {Wagle}\ \emph {et~al.}(2022)\citenamefont {Wagle},
  \citenamefont {Yunes},\ and\ \citenamefont {Silva}}]{Wagle:2021tam}%
  \BibitemOpen
  \bibfield  {author} {\bibinfo {author} {\bibfnamefont {P.}~\bibnamefont
  {Wagle}}, \bibinfo {author} {\bibfnamefont {N.}~\bibnamefont {Yunes}},\ and\
  \bibinfo {author} {\bibfnamefont {H.~O.}\ \bibnamefont {Silva}},\ }\bibfield
  {title} {\bibinfo {title} {{Quasinormal modes of slowly-rotating black holes
  in dynamical Chern-Simons gravity}},\ }\href
  {https://doi.org/10.1103/PhysRevD.105.124003} {\bibfield  {journal} {\bibinfo
   {journal} {Phys. Rev. D}\ }\textbf {\bibinfo {volume} {105}},\ \bibinfo
  {pages} {124003} (\bibinfo {year} {2022})},\ \Eprint
  {https://arxiv.org/abs/2103.09913} {arXiv:2103.09913 [gr-qc]} \BibitemShut
  {NoStop}%
\bibitem [{\citenamefont {Srivastava}\ \emph {et~al.}(2021)\citenamefont
  {Srivastava}, \citenamefont {Chen},\ and\ \citenamefont
  {Shankaranarayanan}}]{Srivastava:2021imr}%
  \BibitemOpen
  \bibfield  {author} {\bibinfo {author} {\bibfnamefont {M.}~\bibnamefont
  {Srivastava}}, \bibinfo {author} {\bibfnamefont {Y.}~\bibnamefont {Chen}},\
  and\ \bibinfo {author} {\bibfnamefont {S.}~\bibnamefont
  {Shankaranarayanan}},\ }\bibfield  {title} {\bibinfo {title} {{Analytical
  computation of quasinormal modes of slowly rotating black holes in dynamical
  Chern-Simons gravity}},\ }\href {https://doi.org/10.1103/PhysRevD.104.064034}
  {\bibfield  {journal} {\bibinfo  {journal} {Phys. Rev. D}\ }\textbf {\bibinfo
  {volume} {104}},\ \bibinfo {pages} {064034} (\bibinfo {year} {2021})},\
  \Eprint {https://arxiv.org/abs/2106.06209} {arXiv:2106.06209 [gr-qc]}
  \BibitemShut {NoStop}%
\bibitem [{\citenamefont {Li}\ \emph {et~al.}(2023)\citenamefont {Li},
  \citenamefont {Wagle}, \citenamefont {Chen},\ and\ \citenamefont
  {Yunes}}]{Li:2022pcy}%
  \BibitemOpen
  \bibfield  {author} {\bibinfo {author} {\bibfnamefont {D.}~\bibnamefont
  {Li}}, \bibinfo {author} {\bibfnamefont {P.}~\bibnamefont {Wagle}}, \bibinfo
  {author} {\bibfnamefont {Y.}~\bibnamefont {Chen}},\ and\ \bibinfo {author}
  {\bibfnamefont {N.}~\bibnamefont {Yunes}},\ }\bibfield  {title} {\bibinfo
  {title} {{Perturbations of Spinning Black Holes beyond General Relativity:
  Modified Teukolsky Equation}},\ }\href
  {https://doi.org/10.1103/PhysRevX.13.021029} {\bibfield  {journal} {\bibinfo
  {journal} {Phys. Rev. X}\ }\textbf {\bibinfo {volume} {13}},\ \bibinfo
  {pages} {021029} (\bibinfo {year} {2023})},\ \Eprint
  {https://arxiv.org/abs/2206.10652} {arXiv:2206.10652 [gr-qc]} \BibitemShut
  {NoStop}%
\bibitem [{\citenamefont {Hussain}\ and\ \citenamefont
  {Zimmerman}(2022)}]{Hussain:2022ins}%
  \BibitemOpen
  \bibfield  {author} {\bibinfo {author} {\bibfnamefont {A.}~\bibnamefont
  {Hussain}}\ and\ \bibinfo {author} {\bibfnamefont {A.}~\bibnamefont
  {Zimmerman}},\ }\bibfield  {title} {\bibinfo {title} {{Approach to computing
  spectral shifts for black holes beyond Kerr}},\ }\href
  {https://doi.org/10.1103/PhysRevD.106.104018} {\bibfield  {journal} {\bibinfo
   {journal} {Phys. Rev. D}\ }\textbf {\bibinfo {volume} {106}},\ \bibinfo
  {pages} {104018} (\bibinfo {year} {2022})},\ \Eprint
  {https://arxiv.org/abs/2206.10653} {arXiv:2206.10653 [gr-qc]} \BibitemShut
  {NoStop}%
\bibitem [{\citenamefont {Cano}\ \emph
  {et~al.}(2023{\natexlab{a}})\citenamefont {Cano}, \citenamefont {Fransen},
  \citenamefont {Hertog},\ and\ \citenamefont {Maenaut}}]{Cano:2023tmv}%
  \BibitemOpen
  \bibfield  {author} {\bibinfo {author} {\bibfnamefont {P.~A.}\ \bibnamefont
  {Cano}}, \bibinfo {author} {\bibfnamefont {K.}~\bibnamefont {Fransen}},
  \bibinfo {author} {\bibfnamefont {T.}~\bibnamefont {Hertog}},\ and\ \bibinfo
  {author} {\bibfnamefont {S.}~\bibnamefont {Maenaut}},\ }\bibfield  {title}
  {\bibinfo {title} {{Universal Teukolsky equations and black hole
  perturbations in higher-derivative gravity}},\ }\href
  {https://doi.org/10.1103/PhysRevD.108.024040} {\bibfield  {journal} {\bibinfo
   {journal} {Phys. Rev. D}\ }\textbf {\bibinfo {volume} {108}},\ \bibinfo
  {pages} {024040} (\bibinfo {year} {2023}{\natexlab{a}})},\ \Eprint
  {https://arxiv.org/abs/2304.02663} {arXiv:2304.02663 [gr-qc]} \BibitemShut
  {NoStop}%
\bibitem [{\citenamefont {Cardoso}\ \emph {et~al.}(2019)\citenamefont
  {Cardoso}, \citenamefont {Kimura}, \citenamefont {Maselli}, \citenamefont
  {Berti}, \citenamefont {Macedo},\ and\ \citenamefont
  {McManus}}]{Cardoso:2019mqo}%
  \BibitemOpen
  \bibfield  {author} {\bibinfo {author} {\bibfnamefont {V.}~\bibnamefont
  {Cardoso}}, \bibinfo {author} {\bibfnamefont {M.}~\bibnamefont {Kimura}},
  \bibinfo {author} {\bibfnamefont {A.}~\bibnamefont {Maselli}}, \bibinfo
  {author} {\bibfnamefont {E.}~\bibnamefont {Berti}}, \bibinfo {author}
  {\bibfnamefont {C.~F.~B.}\ \bibnamefont {Macedo}},\ and\ \bibinfo {author}
  {\bibfnamefont {R.}~\bibnamefont {McManus}},\ }\bibfield  {title} {\bibinfo
  {title} {{Parametrized black hole quasinormal ringdown: Decoupled equations
  for nonrotating black holes}},\ }\href
  {https://doi.org/10.1103/PhysRevD.99.104077} {\bibfield  {journal} {\bibinfo
  {journal} {Phys. Rev. D}\ }\textbf {\bibinfo {volume} {99}},\ \bibinfo
  {pages} {104077} (\bibinfo {year} {2019})},\ \Eprint
  {https://arxiv.org/abs/1901.01265} {arXiv:1901.01265 [gr-qc]} \BibitemShut
  {NoStop}%
\bibitem [{\citenamefont {McManus}\ \emph {et~al.}(2019)\citenamefont
  {McManus}, \citenamefont {Berti}, \citenamefont {Macedo}, \citenamefont
  {Kimura}, \citenamefont {Maselli},\ and\ \citenamefont
  {Cardoso}}]{McManus:2019ulj}%
  \BibitemOpen
  \bibfield  {author} {\bibinfo {author} {\bibfnamefont {R.}~\bibnamefont
  {McManus}}, \bibinfo {author} {\bibfnamefont {E.}~\bibnamefont {Berti}},
  \bibinfo {author} {\bibfnamefont {C.~F.~B.}\ \bibnamefont {Macedo}}, \bibinfo
  {author} {\bibfnamefont {M.}~\bibnamefont {Kimura}}, \bibinfo {author}
  {\bibfnamefont {A.}~\bibnamefont {Maselli}},\ and\ \bibinfo {author}
  {\bibfnamefont {V.}~\bibnamefont {Cardoso}},\ }\bibfield  {title} {\bibinfo
  {title} {{Parametrized black hole quasinormal ringdown. II. Coupled equations
  and quadratic corrections for nonrotating black holes}},\ }\href
  {https://doi.org/10.1103/PhysRevD.100.044061} {\bibfield  {journal} {\bibinfo
   {journal} {Phys. Rev. D}\ }\textbf {\bibinfo {volume} {100}},\ \bibinfo
  {pages} {044061} (\bibinfo {year} {2019})},\ \Eprint
  {https://arxiv.org/abs/1906.05155} {arXiv:1906.05155 [gr-qc]} \BibitemShut
  {NoStop}%
\bibitem [{\citenamefont {V\"olkel}\ \emph
  {et~al.}(2022{\natexlab{a}})\citenamefont {V\"olkel}, \citenamefont
  {Franchini},\ and\ \citenamefont {Barausse}}]{Volkel:2022aca}%
  \BibitemOpen
  \bibfield  {author} {\bibinfo {author} {\bibfnamefont {S.~H.}\ \bibnamefont
  {V\"olkel}}, \bibinfo {author} {\bibfnamefont {N.}~\bibnamefont
  {Franchini}},\ and\ \bibinfo {author} {\bibfnamefont {E.}~\bibnamefont
  {Barausse}},\ }\bibfield  {title} {\bibinfo {title} {{Theory-agnostic
  reconstruction of potential and couplings from quasinormal modes}},\ }\href
  {https://doi.org/10.1103/PhysRevD.105.084046} {\bibfield  {journal} {\bibinfo
   {journal} {Phys. Rev. D}\ }\textbf {\bibinfo {volume} {105}},\ \bibinfo
  {pages} {084046} (\bibinfo {year} {2022}{\natexlab{a}})},\ \Eprint
  {https://arxiv.org/abs/2202.08655} {arXiv:2202.08655 [gr-qc]} \BibitemShut
  {NoStop}%
\bibitem [{\citenamefont {Hirano}\ \emph {et~al.}(2024)\citenamefont {Hirano},
  \citenamefont {Kimura}, \citenamefont {Yamaguchi},\ and\ \citenamefont
  {Zhang}}]{Hirano:2024fgp}%
  \BibitemOpen
  \bibfield  {author} {\bibinfo {author} {\bibfnamefont {S.}~\bibnamefont
  {Hirano}}, \bibinfo {author} {\bibfnamefont {M.}~\bibnamefont {Kimura}},
  \bibinfo {author} {\bibfnamefont {M.}~\bibnamefont {Yamaguchi}},\ and\
  \bibinfo {author} {\bibfnamefont {J.}~\bibnamefont {Zhang}},\ }\bibfield
  {title} {\bibinfo {title} {{Parametrized black hole quasinormal ringdown
  formalism for higher overtones}},\ }\href
  {https://doi.org/10.1103/PhysRevD.110.024015} {\bibfield  {journal} {\bibinfo
   {journal} {Phys. Rev. D}\ }\textbf {\bibinfo {volume} {110}},\ \bibinfo
  {pages} {024015} (\bibinfo {year} {2024})},\ \Eprint
  {https://arxiv.org/abs/2404.09672} {arXiv:2404.09672 [gr-qc]} \BibitemShut
  {NoStop}%
\bibitem [{\citenamefont {V\"olkel}\ \emph
  {et~al.}(2022{\natexlab{b}})\citenamefont {V\"olkel}, \citenamefont
  {Franchini}, \citenamefont {Barausse},\ and\ \citenamefont
  {Berti}}]{Volkel:2022khh}%
  \BibitemOpen
  \bibfield  {author} {\bibinfo {author} {\bibfnamefont {S.~H.}\ \bibnamefont
  {V\"olkel}}, \bibinfo {author} {\bibfnamefont {N.}~\bibnamefont {Franchini}},
  \bibinfo {author} {\bibfnamefont {E.}~\bibnamefont {Barausse}},\ and\
  \bibinfo {author} {\bibfnamefont {E.}~\bibnamefont {Berti}},\ }\bibfield
  {title} {\bibinfo {title} {{Constraining modifications of black hole
  perturbation potentials near the light ring with quasinormal modes}},\ }\href
  {https://doi.org/10.1103/PhysRevD.106.124036} {\bibfield  {journal} {\bibinfo
   {journal} {Phys. Rev. D}\ }\textbf {\bibinfo {volume} {106}},\ \bibinfo
  {pages} {124036} (\bibinfo {year} {2022}{\natexlab{b}})},\ \Eprint
  {https://arxiv.org/abs/2209.10564} {arXiv:2209.10564 [gr-qc]} \BibitemShut
  {NoStop}%
\bibitem [{\citenamefont {Franchini}\ and\ \citenamefont
  {V\"olkel}(2023{\natexlab{b}})}]{Franchini:2022axs}%
  \BibitemOpen
  \bibfield  {author} {\bibinfo {author} {\bibfnamefont {N.}~\bibnamefont
  {Franchini}}\ and\ \bibinfo {author} {\bibfnamefont {S.~H.}\ \bibnamefont
  {V\"olkel}},\ }\bibfield  {title} {\bibinfo {title} {{Parametrized
  quasinormal mode framework for non-Schwarzschild metrics}},\ }\href
  {https://doi.org/10.1103/PhysRevD.107.124063} {\bibfield  {journal} {\bibinfo
   {journal} {Phys. Rev. D}\ }\textbf {\bibinfo {volume} {107}},\ \bibinfo
  {pages} {124063} (\bibinfo {year} {2023}{\natexlab{b}})},\ \Eprint
  {https://arxiv.org/abs/2210.14020} {arXiv:2210.14020 [gr-qc]} \BibitemShut
  {NoStop}%
\bibitem [{\citenamefont {Teukolsky}(1973)}]{Teukolsky:1973ha}%
  \BibitemOpen
  \bibfield  {author} {\bibinfo {author} {\bibfnamefont {S.~A.}\ \bibnamefont
  {Teukolsky}},\ }\bibfield  {title} {\bibinfo {title} {{Perturbations of a
  rotating black hole. 1. Fundamental equations for gravitational
  electromagnetic and neutrino field perturbations}},\ }\href
  {https://doi.org/10.1086/152444} {\bibfield  {journal} {\bibinfo  {journal}
  {Astrophys. J.}\ }\textbf {\bibinfo {volume} {185}},\ \bibinfo {pages} {635}
  (\bibinfo {year} {1973})}\BibitemShut {NoStop}%
\bibitem [{\citenamefont {Ghosh}\ \emph {et~al.}(2023)\citenamefont {Ghosh},
  \citenamefont {Franchini}, \citenamefont {V\"olkel},\ and\ \citenamefont
  {Barausse}}]{Ghosh:2023etd}%
  \BibitemOpen
  \bibfield  {author} {\bibinfo {author} {\bibfnamefont {R.}~\bibnamefont
  {Ghosh}}, \bibinfo {author} {\bibfnamefont {N.}~\bibnamefont {Franchini}},
  \bibinfo {author} {\bibfnamefont {S.~H.}\ \bibnamefont {V\"olkel}},\ and\
  \bibinfo {author} {\bibfnamefont {E.}~\bibnamefont {Barausse}},\ }\bibfield
  {title} {\bibinfo {title} {{Quasinormal modes of nonseparable perturbation
  equations: The scalar non-Kerr case}},\ }\href
  {https://doi.org/10.1103/PhysRevD.108.024038} {\bibfield  {journal} {\bibinfo
   {journal} {Phys. Rev. D}\ }\textbf {\bibinfo {volume} {108}},\ \bibinfo
  {pages} {024038} (\bibinfo {year} {2023})},\ \Eprint
  {https://arxiv.org/abs/2303.00088} {arXiv:2303.00088 [gr-qc]} \BibitemShut
  {NoStop}%
\bibitem [{\citenamefont {Kimura}(2020)}]{Kimura:2020mrh}%
  \BibitemOpen
  \bibfield  {author} {\bibinfo {author} {\bibfnamefont {M.}~\bibnamefont
  {Kimura}},\ }\bibfield  {title} {\bibinfo {title} {{Note on the parametrized
  black hole quasinormal ringdown formalism}},\ }\href
  {https://doi.org/10.1103/PhysRevD.101.064031} {\bibfield  {journal} {\bibinfo
   {journal} {Phys. Rev. D}\ }\textbf {\bibinfo {volume} {101}},\ \bibinfo
  {pages} {064031} (\bibinfo {year} {2020})},\ \Eprint
  {https://arxiv.org/abs/2001.09613} {arXiv:2001.09613 [gr-qc]} \BibitemShut
  {NoStop}%
\bibitem [{\citenamefont {Cano}\ \emph {et~al.}(2024)\citenamefont {Cano},
  \citenamefont {Capuano}, \citenamefont {Franchini}, \citenamefont {Maenaut},\
  and\ \citenamefont {V\"olkel}}]{Cano:2024ezp}%
  \BibitemOpen
  \bibfield  {author} {\bibinfo {author} {\bibfnamefont {P.~A.}\ \bibnamefont
  {Cano}}, \bibinfo {author} {\bibfnamefont {L.}~\bibnamefont {Capuano}},
  \bibinfo {author} {\bibfnamefont {N.}~\bibnamefont {Franchini}}, \bibinfo
  {author} {\bibfnamefont {S.}~\bibnamefont {Maenaut}},\ and\ \bibinfo {author}
  {\bibfnamefont {S.~H.}\ \bibnamefont {V\"olkel}},\ }\bibfield  {title}
  {\bibinfo {title} {{Higher-derivative corrections to the Kerr quasinormal
  mode spectrum}},\ }\href@noop {} {\  (\bibinfo {year} {2024})},\ \Eprint
  {https://arxiv.org/abs/2409.04517} {arXiv:2409.04517 [gr-qc]} \BibitemShut
  {NoStop}%
\bibitem [{\citenamefont {Cano}\ \emph
  {et~al.}(2023{\natexlab{b}})\citenamefont {Cano}, \citenamefont {Fransen},
  \citenamefont {Hertog},\ and\ \citenamefont {Maenaut}}]{Cano:2023jbk}%
  \BibitemOpen
  \bibfield  {author} {\bibinfo {author} {\bibfnamefont {P.~A.}\ \bibnamefont
  {Cano}}, \bibinfo {author} {\bibfnamefont {K.}~\bibnamefont {Fransen}},
  \bibinfo {author} {\bibfnamefont {T.}~\bibnamefont {Hertog}},\ and\ \bibinfo
  {author} {\bibfnamefont {S.}~\bibnamefont {Maenaut}},\ }\bibfield  {title}
  {\bibinfo {title} {{Quasinormal modes of rotating black holes in
  higher-derivative gravity}},\ }\href
  {https://doi.org/10.1103/PhysRevD.108.124032} {\bibfield  {journal} {\bibinfo
   {journal} {Phys. Rev. D}\ }\textbf {\bibinfo {volume} {108}},\ \bibinfo
  {pages} {124032} (\bibinfo {year} {2023}{\natexlab{b}})},\ \Eprint
  {https://arxiv.org/abs/2307.07431} {arXiv:2307.07431 [gr-qc]} \BibitemShut
  {NoStop}%
\bibitem [{\citenamefont {Leaver}(1985)}]{Leaver:1985ax}%
  \BibitemOpen
  \bibfield  {author} {\bibinfo {author} {\bibfnamefont {E.~W.}\ \bibnamefont
  {Leaver}},\ }\bibfield  {title} {\bibinfo {title} {{An Analytic
  representation for the quasi normal modes of Kerr black holes}},\ }\href
  {https://doi.org/10.1098/rspa.1985.0119} {\bibfield  {journal} {\bibinfo
  {journal} {Proc. Roy. Soc. Lond. A}\ }\textbf {\bibinfo {volume} {402}},\
  \bibinfo {pages} {285} (\bibinfo {year} {1985})}\BibitemShut {NoStop}%
\bibitem [{\citenamefont {github}(2024)}]{github}%
  \BibitemOpen
  \bibfield  {author} {\bibinfo {author} {\bibnamefont {github}},\ }\href
  {https://github.com/sebastianvoelkel/parametrized_qnm_framework} {\bibinfo
  {title} {{parametrized\_qnm\_framework}}} (\bibinfo {year}
  {2024})\BibitemShut {NoStop}%
\bibitem [{\citenamefont {Berti}\ \emph {et~al.}(2006)\citenamefont {Berti},
  \citenamefont {Cardoso},\ and\ \citenamefont {Will}}]{Berti:2005ys}%
  \BibitemOpen
  \bibfield  {author} {\bibinfo {author} {\bibfnamefont {E.}~\bibnamefont
  {Berti}}, \bibinfo {author} {\bibfnamefont {V.}~\bibnamefont {Cardoso}},\
  and\ \bibinfo {author} {\bibfnamefont {C.~M.}\ \bibnamefont {Will}},\
  }\bibfield  {title} {\bibinfo {title} {{On gravitational-wave spectroscopy of
  massive black holes with the space interferometer LISA}},\ }\href
  {https://doi.org/10.1103/PhysRevD.73.064030} {\bibfield  {journal} {\bibinfo
  {journal} {Phys. Rev. D}\ }\textbf {\bibinfo {volume} {73}},\ \bibinfo
  {pages} {064030} (\bibinfo {year} {2006})},\ \Eprint
  {https://arxiv.org/abs/gr-qc/0512160} {arXiv:gr-qc/0512160} \BibitemShut
  {NoStop}%
\bibitem [{\citenamefont {Bardeen}\ and\ \citenamefont
  {Press}(1973)}]{Bardeen:1973xb}%
  \BibitemOpen
  \bibfield  {author} {\bibinfo {author} {\bibfnamefont {J.~M.}\ \bibnamefont
  {Bardeen}}\ and\ \bibinfo {author} {\bibfnamefont {W.~H.}\ \bibnamefont
  {Press}},\ }\bibfield  {title} {\bibinfo {title} {{Radiation fields in the
  schwarzschild background}},\ }\href {https://doi.org/10.1063/1.1666175}
  {\bibfield  {journal} {\bibinfo  {journal} {J. Math. Phys.}\ }\textbf
  {\bibinfo {volume} {14}},\ \bibinfo {pages} {7} (\bibinfo {year}
  {1973})}\BibitemShut {NoStop}%
\bibitem [{\citenamefont {Chandrasekhar}(1975)}]{Chandrasekhar:1975nkd}%
  \BibitemOpen
  \bibfield  {author} {\bibinfo {author} {\bibfnamefont {S.}~\bibnamefont
  {Chandrasekhar}},\ }\bibfield  {title} {\bibinfo {title} {{On the equations
  governing the perturbations of the Schwarzschild black hole}},\ }\href
  {https://doi.org/10.1098/rspa.1975.0066} {\bibfield  {journal} {\bibinfo
  {journal} {Proc. Roy. Soc. Lond. A}\ }\textbf {\bibinfo {volume} {343}},\
  \bibinfo {pages} {289} (\bibinfo {year} {1975})}\BibitemShut {NoStop}%
\bibitem [{\citenamefont {Zouros}\ and\ \citenamefont
  {Eardley}(1979)}]{Zouros:1979iw}%
  \BibitemOpen
  \bibfield  {author} {\bibinfo {author} {\bibfnamefont {T.~J.~M.}\
  \bibnamefont {Zouros}}\ and\ \bibinfo {author} {\bibfnamefont {D.~M.}\
  \bibnamefont {Eardley}},\ }\bibfield  {title} {\bibinfo {title}
  {{Instabilities of massive scalar perturbations of a rotating black hole}},\
  }\href {https://doi.org/10.1016/0003-4916(79)90237-9} {\bibfield  {journal}
  {\bibinfo  {journal} {Annals Phys.}\ }\textbf {\bibinfo {volume} {118}},\
  \bibinfo {pages} {139} (\bibinfo {year} {1979})}\BibitemShut {NoStop}%
\bibitem [{\citenamefont {Detweiler}(1980)}]{Detweiler:1980uk}%
  \BibitemOpen
  \bibfield  {author} {\bibinfo {author} {\bibfnamefont {S.~L.}\ \bibnamefont
  {Detweiler}},\ }\bibfield  {title} {\bibinfo {title} {{Klein-Gordon equation
  and rotating black holes}},\ }\href
  {https://doi.org/10.1103/PhysRevD.22.2323} {\bibfield  {journal} {\bibinfo
  {journal} {Phys. Rev. D}\ }\textbf {\bibinfo {volume} {22}},\ \bibinfo
  {pages} {2323} (\bibinfo {year} {1980})}\BibitemShut {NoStop}%
\bibitem [{\citenamefont {Dolan}(2007)}]{Dolan:2007mj}%
  \BibitemOpen
  \bibfield  {author} {\bibinfo {author} {\bibfnamefont {S.~R.}\ \bibnamefont
  {Dolan}},\ }\bibfield  {title} {\bibinfo {title} {{Instability of the massive
  Klein-Gordon field on the Kerr spacetime}},\ }\href
  {https://doi.org/10.1103/PhysRevD.76.084001} {\bibfield  {journal} {\bibinfo
  {journal} {Phys. Rev. D}\ }\textbf {\bibinfo {volume} {76}},\ \bibinfo
  {pages} {084001} (\bibinfo {year} {2007})},\ \Eprint
  {https://arxiv.org/abs/0705.2880} {arXiv:0705.2880 [gr-qc]} \BibitemShut
  {NoStop}%
\bibitem [{\citenamefont {Dias}\ \emph {et~al.}(2022)\citenamefont {Dias},
  \citenamefont {Godazgar}, \citenamefont {Santos}, \citenamefont {Carullo},
  \citenamefont {Del~Pozzo},\ and\ \citenamefont {Laghi}}]{Dias:2021yju}%
  \BibitemOpen
  \bibfield  {author} {\bibinfo {author} {\bibfnamefont {O.~J.~C.}\
  \bibnamefont {Dias}}, \bibinfo {author} {\bibfnamefont {M.}~\bibnamefont
  {Godazgar}}, \bibinfo {author} {\bibfnamefont {J.~E.}\ \bibnamefont
  {Santos}}, \bibinfo {author} {\bibfnamefont {G.}~\bibnamefont {Carullo}},
  \bibinfo {author} {\bibfnamefont {W.}~\bibnamefont {Del~Pozzo}},\ and\
  \bibinfo {author} {\bibfnamefont {D.}~\bibnamefont {Laghi}},\ }\bibfield
  {title} {\bibinfo {title} {{Eigenvalue repulsions in the quasinormal spectra
  of the Kerr-Newman black hole}},\ }\href
  {https://doi.org/10.1103/PhysRevD.105.084044} {\bibfield  {journal} {\bibinfo
   {journal} {Phys. Rev. D}\ }\textbf {\bibinfo {volume} {105}},\ \bibinfo
  {pages} {084044} (\bibinfo {year} {2022})},\ \Eprint
  {https://arxiv.org/abs/2109.13949} {arXiv:2109.13949 [gr-qc]} \BibitemShut
  {NoStop}%
\bibitem [{\citenamefont {Chandrasekhar}(1984)}]{Chandrasekhar:1984siy}%
  \BibitemOpen
  \bibfield  {author} {\bibinfo {author} {\bibfnamefont {S.}~\bibnamefont
  {Chandrasekhar}},\ }\bibfield  {title} {\bibinfo {title} {{The Mathematical
  Theory of Black Holes}},\ }\href
  {https://doi.org/10.1007/978-94-009-6469-3_2} {\bibfield  {journal} {\bibinfo
   {journal} {Fundam. Theor. Phys.}\ }\textbf {\bibinfo {volume} {9}},\
  \bibinfo {pages} {5} (\bibinfo {year} {1984})}\BibitemShut {NoStop}%
\bibitem [{\citenamefont {Dudley}\ and\ \citenamefont
  {Finley}(1977)}]{Dudley:1977zz}%
  \BibitemOpen
  \bibfield  {author} {\bibinfo {author} {\bibfnamefont {A.~L.}\ \bibnamefont
  {Dudley}}\ and\ \bibinfo {author} {\bibfnamefont {J.~D.}\ \bibnamefont
  {Finley}},\ }\bibfield  {title} {\bibinfo {title} {{Separation of Wave
  Equations for Perturbations of General Type-D Space-Times}},\ }\href
  {https://doi.org/10.1103/PhysRevLett.38.1505} {\bibfield  {journal} {\bibinfo
   {journal} {Phys. Rev. Lett.}\ }\textbf {\bibinfo {volume} {38}},\ \bibinfo
  {pages} {1505} (\bibinfo {year} {1977})}\BibitemShut {NoStop}%
\bibitem [{\citenamefont {Dudley}\ and\ \citenamefont
  {Finley}(1979)}]{Dudley:1978vd}%
  \BibitemOpen
  \bibfield  {author} {\bibinfo {author} {\bibfnamefont {A.~L.}\ \bibnamefont
  {Dudley}}\ and\ \bibinfo {author} {\bibfnamefont {J.~D.}\ \bibnamefont
  {Finley}, \bibfnamefont {III}},\ }\bibfield  {title} {\bibinfo {title}
  {{Covariant Perturbed Wave Equations in Arbitrary Type $D$ Backgrounds}},\
  }\href {https://doi.org/10.1063/1.524064} {\bibfield  {journal} {\bibinfo
  {journal} {J. Math. Phys.}\ }\textbf {\bibinfo {volume} {20}},\ \bibinfo
  {pages} {311} (\bibinfo {year} {1979})}\BibitemShut {NoStop}%
\bibitem [{\citenamefont {Berti}\ and\ \citenamefont
  {Kokkotas}(2005)}]{Berti:2005eb}%
  \BibitemOpen
  \bibfield  {author} {\bibinfo {author} {\bibfnamefont {E.}~\bibnamefont
  {Berti}}\ and\ \bibinfo {author} {\bibfnamefont {K.~D.}\ \bibnamefont
  {Kokkotas}},\ }\bibfield  {title} {\bibinfo {title} {{Quasinormal modes of
  Kerr-Newman black holes: Coupling of electromagnetic and gravitational
  perturbations}},\ }\href {https://doi.org/10.1103/PhysRevD.71.124008}
  {\bibfield  {journal} {\bibinfo  {journal} {Phys. Rev. D}\ }\textbf {\bibinfo
  {volume} {71}},\ \bibinfo {pages} {124008} (\bibinfo {year} {2005})},\
  \Eprint {https://arxiv.org/abs/gr-qc/0502065} {arXiv:gr-qc/0502065}
  \BibitemShut {NoStop}%
\bibitem [{\citenamefont {Wagle}\ \emph {et~al.}(2024)\citenamefont {Wagle},
  \citenamefont {Li}, \citenamefont {Chen},\ and\ \citenamefont
  {Yunes}}]{Wagle:2023fwl}%
  \BibitemOpen
  \bibfield  {author} {\bibinfo {author} {\bibfnamefont {P.}~\bibnamefont
  {Wagle}}, \bibinfo {author} {\bibfnamefont {D.}~\bibnamefont {Li}}, \bibinfo
  {author} {\bibfnamefont {Y.}~\bibnamefont {Chen}},\ and\ \bibinfo {author}
  {\bibfnamefont {N.}~\bibnamefont {Yunes}},\ }\bibfield  {title} {\bibinfo
  {title} {{Perturbations of spinning black holes in dynamical Chern-Simons
  gravity: Slow rotation equations}},\ }\href
  {https://doi.org/10.1103/PhysRevD.109.104029} {\bibfield  {journal} {\bibinfo
   {journal} {Phys. Rev. D}\ }\textbf {\bibinfo {volume} {109}},\ \bibinfo
  {pages} {104029} (\bibinfo {year} {2024})},\ \Eprint
  {https://arxiv.org/abs/2311.07706} {arXiv:2311.07706 [gr-qc]} \BibitemShut
  {NoStop}%
\bibitem [{\citenamefont {Chung}\ \emph {et~al.}(2023)\citenamefont {Chung},
  \citenamefont {Wagle},\ and\ \citenamefont {Yunes}}]{Chung:2023zdq}%
  \BibitemOpen
  \bibfield  {author} {\bibinfo {author} {\bibfnamefont {A.~K.-W.}\
  \bibnamefont {Chung}}, \bibinfo {author} {\bibfnamefont {P.}~\bibnamefont
  {Wagle}},\ and\ \bibinfo {author} {\bibfnamefont {N.}~\bibnamefont {Yunes}},\
  }\bibfield  {title} {\bibinfo {title} {{Spectral method for the gravitational
  perturbations of black holes: Schwarzschild background case}},\ }\href
  {https://doi.org/10.1103/PhysRevD.107.124032} {\bibfield  {journal} {\bibinfo
   {journal} {Phys. Rev. D}\ }\textbf {\bibinfo {volume} {107}},\ \bibinfo
  {pages} {124032} (\bibinfo {year} {2023})},\ \Eprint
  {https://arxiv.org/abs/2302.11624} {arXiv:2302.11624 [gr-qc]} \BibitemShut
  {NoStop}%
\bibitem [{\citenamefont {Chung}\ \emph {et~al.}(2024)\citenamefont {Chung},
  \citenamefont {Wagle},\ and\ \citenamefont {Yunes}}]{Chung:2023wkd}%
  \BibitemOpen
  \bibfield  {author} {\bibinfo {author} {\bibfnamefont {A.~K.-W.}\
  \bibnamefont {Chung}}, \bibinfo {author} {\bibfnamefont {P.}~\bibnamefont
  {Wagle}},\ and\ \bibinfo {author} {\bibfnamefont {N.}~\bibnamefont {Yunes}},\
  }\bibfield  {title} {\bibinfo {title} {{Spectral method for metric
  perturbations of black holes: Kerr background case in general relativity}},\
  }\href {https://doi.org/10.1103/PhysRevD.109.044072} {\bibfield  {journal}
  {\bibinfo  {journal} {Phys. Rev. D}\ }\textbf {\bibinfo {volume} {109}},\
  \bibinfo {pages} {044072} (\bibinfo {year} {2024})},\ \Eprint
  {https://arxiv.org/abs/2312.08435} {arXiv:2312.08435 [gr-qc]} \BibitemShut
  {NoStop}%
\bibitem [{\citenamefont {Bl\'azquez-Salcedo}\ \emph
  {et~al.}(2024)\citenamefont {Bl\'azquez-Salcedo}, \citenamefont {Khoo},
  \citenamefont {Kunz},\ and\ \citenamefont
  {Gonz\'alez-Romero}}]{Blazquez-Salcedo:2023hwg}%
  \BibitemOpen
  \bibfield  {author} {\bibinfo {author} {\bibfnamefont {J.~L.}\ \bibnamefont
  {Bl\'azquez-Salcedo}}, \bibinfo {author} {\bibfnamefont {F.~S.}\ \bibnamefont
  {Khoo}}, \bibinfo {author} {\bibfnamefont {J.}~\bibnamefont {Kunz}},\ and\
  \bibinfo {author} {\bibfnamefont {L.~M.}\ \bibnamefont {Gonz\'alez-Romero}},\
  }\bibfield  {title} {\bibinfo {title} {{Quasinormal modes of Kerr black holes
  using a spectral decomposition of the metric perturbations}},\ }\href
  {https://doi.org/10.1103/PhysRevD.109.064028} {\bibfield  {journal} {\bibinfo
   {journal} {Phys. Rev. D}\ }\textbf {\bibinfo {volume} {109}},\ \bibinfo
  {pages} {064028} (\bibinfo {year} {2024})},\ \Eprint
  {https://arxiv.org/abs/2312.10754} {arXiv:2312.10754 [gr-qc]} \BibitemShut
  {NoStop}%
\bibitem [{\citenamefont {Chung}\ and\ \citenamefont
  {Yunes}(2024{\natexlab{a}})}]{Chung:2024ira}%
  \BibitemOpen
  \bibfield  {author} {\bibinfo {author} {\bibfnamefont {A.~K.-W.}\
  \bibnamefont {Chung}}\ and\ \bibinfo {author} {\bibfnamefont
  {N.}~\bibnamefont {Yunes}},\ }\bibfield  {title} {\bibinfo {title} {{Ringing
  out General Relativity: Quasi-normal mode frequencies for black holes of any
  spin in modified gravity}},\ }\href@noop {} {\  (\bibinfo {year}
  {2024}{\natexlab{a}})},\ \Eprint {https://arxiv.org/abs/2405.12280}
  {arXiv:2405.12280 [gr-qc]} \BibitemShut {NoStop}%
\bibitem [{\citenamefont {Chung}\ and\ \citenamefont
  {Yunes}(2024{\natexlab{b}})}]{Chung:2024vaf}%
  \BibitemOpen
  \bibfield  {author} {\bibinfo {author} {\bibfnamefont {A.~K.-W.}\
  \bibnamefont {Chung}}\ and\ \bibinfo {author} {\bibfnamefont
  {N.}~\bibnamefont {Yunes}},\ }\bibfield  {title} {\bibinfo {title}
  {{Quasi-normal mode frequencies and gravitational perturbations of black
  holes with any subextremal spin in modified gravity through METRICS: the
  scalar-Gauss-Bonnet gravity case}},\ }\href@noop {} {\  (\bibinfo {year}
  {2024}{\natexlab{b}})},\ \Eprint {https://arxiv.org/abs/2406.11986}
  {arXiv:2406.11986 [gr-qc]} \BibitemShut {NoStop}%
\bibitem [{\citenamefont {Mitman}\ \emph {et~al.}(2023)\citenamefont {Mitman}
  \emph {et~al.}}]{Mitman:2022qdl}%
  \BibitemOpen
  \bibfield  {author} {\bibinfo {author} {\bibfnamefont {K.}~\bibnamefont
  {Mitman}} \emph {et~al.},\ }\bibfield  {title} {\bibinfo {title}
  {{Nonlinearities in Black Hole Ringdowns}},\ }\href
  {https://doi.org/10.1103/PhysRevLett.130.081402} {\bibfield  {journal}
  {\bibinfo  {journal} {Phys. Rev. Lett.}\ }\textbf {\bibinfo {volume} {130}},\
  \bibinfo {pages} {081402} (\bibinfo {year} {2023})},\ \Eprint
  {https://arxiv.org/abs/2208.07380} {arXiv:2208.07380 [gr-qc]} \BibitemShut
  {NoStop}%
\bibitem [{\citenamefont {Cheung}\ \emph {et~al.}(2023)\citenamefont {Cheung}
  \emph {et~al.}}]{Cheung:2022rbm}%
  \BibitemOpen
  \bibfield  {author} {\bibinfo {author} {\bibfnamefont {M.~H.-Y.}\
  \bibnamefont {Cheung}} \emph {et~al.},\ }\bibfield  {title} {\bibinfo {title}
  {{Nonlinear Effects in Black Hole Ringdown}},\ }\href
  {https://doi.org/10.1103/PhysRevLett.130.081401} {\bibfield  {journal}
  {\bibinfo  {journal} {Phys. Rev. Lett.}\ }\textbf {\bibinfo {volume} {130}},\
  \bibinfo {pages} {081401} (\bibinfo {year} {2023})},\ \Eprint
  {https://arxiv.org/abs/2208.07374} {arXiv:2208.07374 [gr-qc]} \BibitemShut
  {NoStop}%
\bibitem [{\citenamefont {Yi}\ \emph {et~al.}(2024)\citenamefont {Yi},
  \citenamefont {Kuntz}, \citenamefont {Barausse}, \citenamefont {Berti},
  \citenamefont {Cheung}, \citenamefont {Kritos},\ and\ \citenamefont
  {Maselli}}]{Yi:2024elj}%
  \BibitemOpen
  \bibfield  {author} {\bibinfo {author} {\bibfnamefont {S.}~\bibnamefont
  {Yi}}, \bibinfo {author} {\bibfnamefont {A.}~\bibnamefont {Kuntz}}, \bibinfo
  {author} {\bibfnamefont {E.}~\bibnamefont {Barausse}}, \bibinfo {author}
  {\bibfnamefont {E.}~\bibnamefont {Berti}}, \bibinfo {author} {\bibfnamefont
  {M.~H.-Y.}\ \bibnamefont {Cheung}}, \bibinfo {author} {\bibfnamefont
  {K.}~\bibnamefont {Kritos}},\ and\ \bibinfo {author} {\bibfnamefont
  {A.}~\bibnamefont {Maselli}},\ }\bibfield  {title} {\bibinfo {title}
  {{Nonlinear quasinormal mode detectability with next-generation gravitational
  wave detectors}},\ }\href {https://doi.org/10.1103/PhysRevD.109.124029}
  {\bibfield  {journal} {\bibinfo  {journal} {Phys. Rev. D}\ }\textbf {\bibinfo
  {volume} {109}},\ \bibinfo {pages} {124029} (\bibinfo {year} {2024})},\
  \Eprint {https://arxiv.org/abs/2403.09767} {arXiv:2403.09767 [gr-qc]}
  \BibitemShut {NoStop}%
\bibitem [{\citenamefont {Nollert}(1993)}]{Nollert:1993zz}%
  \BibitemOpen
  \bibfield  {author} {\bibinfo {author} {\bibfnamefont {H.~P.}\ \bibnamefont
  {Nollert}},\ }\bibfield  {title} {\bibinfo {title} {{Quasinormal modes of
  Schwarzschild black holes: The determination of quasinormal frequencies with
  very large imaginary parts}},\ }\href
  {https://doi.org/10.1103/PhysRevD.47.5253} {\bibfield  {journal} {\bibinfo
  {journal} {Phys. Rev. D}\ }\textbf {\bibinfo {volume} {47}},\ \bibinfo
  {pages} {5253} (\bibinfo {year} {1993})}\BibitemShut {NoStop}%
\end{thebibliography}%

\end{document}